\documentclass[10pt,twocolumn,letterpaper]{article}

\usepackage{cvpr}              % To produce the CAMERA-READY version
\usepackage[pagebackref,breaklinks,colorlinks,citecolor=cvprblue]{hyperref}
\usepackage{framed,multirow}

\usepackage{amsmath,amssymb,amsfonts}
\usepackage{latexsym}
\usepackage{algorithm, algorithmic}
\usepackage{graphicx}
\usepackage{textcomp}
\usepackage{newfloat}
\usepackage{caption}
\usepackage{listings}
\usepackage{subfigure}
\usepackage{booktabs}
\usepackage{textcomp}
\usepackage{array, multirow, bigdelim, makecell}

\newcommand{\rev}[1]{\textcolor{black}{#1}}
\newcommand{\revfinal}[1]{\textcolor{black}{#1}}

\usepackage{url}
\usepackage{xcolor}
\usepackage{hyperref}

\definecolor{newcolor}{rgb}{.8,.349,.1}
\definecolor{cvprblue}{rgb}{0.21,0.49,0.74}

\def\paperID{*****} % *** Enter the Paper ID here
\def\confName{CVPR}
\def\confYear{2024}

\title{Unsupervised anomaly localization in high-resolution breast scans using deep pluralistic image completion}

\author{%
  Nicholas Konz$^{1}$ \quad Haoyu Dong$^{1}$ \quad Maciej A. Mazurowski$^{1,2,3,4}$\\
  $^{1}$ Department of Electrical and Computer Engineering $^{2}$ Department of Computer Science \\
  $^{3}$ Department of Radiology \quad $^{4}$ Department of Biostatistics \& Bioinformatics \\
  Duke University, NC, USA\\
  {\tt\small \{nicholas.konz, haoyu.dong151, maciej.mazurowski\}@duke.edu} \\
}

\begin{document}
\maketitle

\begin{abstract}
    %%%
    Automated tumor detection in Digital Breast Tomosynthesis (DBT) is a difficult task due to natural tumor rarity, breast tissue variability, and high resolution. 
    Given the scarcity of abnormal images and the abundance of normal images for this problem, an anomaly detection/localization approach could be well-suited.
    However, most anomaly localization research in machine learning focuses on non-medical datasets, and we find that these methods fall short when adapted to medical imaging datasets. 
    The problem is alleviated when we solve the task from the image completion perspective, in which the presence of anomalies can be indicated by a discrepancy between the original appearance and its auto-completion conditioned on the surroundings. 
    However, there are often many valid normal completions given the same surroundings, especially in the DBT dataset, making this evaluation criterion less precise. 
    To address such an issue, we consider \textit{pluralistic image completion} by exploring the distribution of possible completions instead of generating fixed predictions.
    This is achieved through our novel application of spatial dropout on the completion network during inference time only, which requires no additional training cost and is effective at generating diverse completions.
    We further propose \textit{minimum completion distance} (MCD), a new metric for detecting anomalies, thanks to these stochastic completions.
    We provide theoretical as well as empirical support for the superiority over existing methods of using the proposed method for anomaly localization. On the DBT dataset, our model outperforms other state-of-the-art methods by at least 10\% AUROC for pixel-level detection.
    
    %%%%
\end{abstract}

\section{Introduction}
\label{sec:intro}

Anomaly detection (AD) refers to the task of detecting patterns in data that are not present in normal data. It is an important and safety-critical task in medical imaging and many other fields. 
In many situations, little or no anomalous data is available, making it crucial to develop methods that can perform AD using \textit{only normal data} for training, \rev{a task known as \textit{unsupervised anomaly detection} \cite{bergmann2019mvtec}. This is because traditional \textit{supervised} computer vision models requires large amounts of both normal and anomalous data for training, making them not applicable to this scenario.}
This research direction is especially important in the medical imaging field where it is often resource-intensive to acquire new data. Such methods are referred to as \textit{unsupervised} or \textit{self-supervised} learning methods. 

In this manuscript, we consider the realistic case of Digital Breast Tomosynthesis (DBT) data, a relatively new breast cancer screening modality that has gained traction in recent years. It is difficult to develop AD methods for these images, due to the high rarity of cancer cases \rev{and the natural anatomical variability seen in both healthy and cancerous cases. The high resolution of DBT poses an additional challenge for tumor detection methods because the images cannot always be downsampled to a lower resolution without losing the fine-grained anatomical detail present in breast tissue that may be necessary for accurate tumor detection.}

\rev{Indeed, we find that standard deep learning-based AD methods, which perform well on non medical-image domains, have poor performance on DBT scans. Deep methods are vulnerable to (1) the high visual similarity of certain normal and cancerous DBT images that leads to images of different classes appearing very similar, and (2) the aforementioned high resolution problem, which are both issues that are less present in the natural image datasets that standard deep AD methods are evaluated on (\eg, \cite{bergmann2019mvtec}).}
This motivates our advanced unsupervised image anomaly detection method, which solves the problem from a different perspective.

An intuitive way of thinking about an anomalous image is that the content in the image is unexpected, given knowledge of what normal data looks like.
This intuition can be implemented in the unsupervised or self-supervised regime, as it does not require any explicit knowledge about anomalous data. In particular, we solved this problem through image completion beginning with our earlier work of \cite{swiecicki2021generative}, \ie, if some region of an image is removed and a completion network is asked to ``fill in'' a normal prediction given the surroundings, and the predicted region and the original region are different, then that region can be considered anomalous.

However, a shortcoming of this approach is that 
the output completion, despite being realistic, is fixed for a given input.
False positives can occur if only a single possible (\textit{deterministic}) normal completion is predicted by the network, given that there can be various valid completions for a region. 
Many masked images theoretically have a multimodal distribution of possible completions, so a completely anomaly-free ground truth region could be distinctively different from the completion that the network happens to output. 
In other words, any dissimilarity of the original image to just one of the possible predictions is an imprecise measure of abnormality. Moreover, the presence of multiple valid completions is especially prominent in data with high semantic variability, such as breast tissue scans.

To remedy this problem, our approach uses a \textit{pluralistic} image completion network to sample from the distribution of possible normal completions to compare to the ground truth, which we achieve using a novel and simple application of spatial (channel-wise) dropout layers to a pretrained image completion network. Even if certain surroundings of a normal ground truth have a high number of semantically distinct valid normal completions, we are guaranteed to eventually sample a completion that is similar to the ground truth, provided that the pluralistic network is a strong approximation of the true distribution of valid normal completions. However, if the ground truth is anomalous, then it is very unlikely that \textit{any} valid normal completion is similar to it, because the two samples are from fundamentally distinct distributions. 

Following this observation, we expect that given a large sample of normal completions, if the ground truth is normal, the distance of the closest completion to the ground truth will be greater if the ground truth is anomalous than if it is normal. We can quantify this idea by taking the minimum of all of the distances from each completion to the ground truth, and hypothesizing that this minimum distance will generally be greater for anomalous ground truths than for normal ground truths. From these ideas we propose a new anomaly score metric: \textit{minimum completion distance}, or \textit{MCD}.
We have shown both theoretically and empirically that it is a more faithful measure of abnormality.

Given a 2D slice of a pseudo-3D DBT scan volume, our method works by sampling multiple completions of successive patches on a ``sliding'' raster window on the slice/image. We can detect anomalies within the spatial location of each patch using our MCD metric, which analyzes how similar the ground truth of the completion region is to the sampled completions; if the ground truth is sufficiently different from the sampled completions, we assume that the region contains an anomaly. This procedure is performed on many overlapping patches that cover the entire image, so that a full anomaly heatmap can be generated at the end using the spatially-oriented anomaly scores of each patch. \rev{We perform patch-level anomaly detections in parallel, along both the number of completions to sample per patch, and the number of patches to complete.} Our overall method is named \textit{PICARD}, or Pluralistic Image Completion for Anomalous Representation Detection. \revfinal{We provide Python/PyTorch code for our method at \url{https://github.com/mazurowski-lab/picard}}.

\paragraph{Novel Contributions}
In summary, our contributions are the following:
\begin{enumerate}
\item We introduce a novel anomaly localization model that uses channel-wise dropout on image patches to rapidly sample \textit{pluralistic} completions of patches in order to localize anomalies on the image. 
\item We propose a novel evaluation metric, MCD, for completion similarity assessment and anomaly scoring. We provide a thorough analysis of the effectiveness of this metric.
\item By adopting existing state-of-the-art methods that aim for natural / low-resolution images, we build an anomaly localization performance benchmark on the challenging DBT dataset, in which our method outperforms these methods by a large margin. This benchmark also serves as a foundation for future works.
\end{enumerate}

The rest of this manuscript is organized as follows: In Section \ref{sec:rel_works}, we explore related works. In Section \ref{sec:methods} we mathematically analyze the effectiveness of MCD, and our method used to achieve \textit{pluralistic} image completion. In Sections \ref{sec:data} and \ref{sec:experiments} we present our target dataset and experimental results, respectively. Finally, in Section \ref{sec:disc} we discuss our findings and outline future research directions, and in Section \ref{sec:concl} we summarize our conclusions.

\section{Related Works}
\label{sec:rel_works}
\subsection*{Anomaly Localization with Self-Supervised Learning}

The tasks of anomaly \textit{detection} (AD), \ie, the classification of entire images as being either normal or anomalous, and anomaly \textit{localization}/segmentation (AL), \ie, the spatial segmentation of anomalies within images, have received considerable attention within the fields of machine learning and deep learning in particular. Many AD works exist \cite{choi2018waic,ren2019likelihood,grathwohl2019your,nalisnick2018deep,nalisnick2019detecting,serra2019input,du2019implicit,schlegl2019f,choi2019novelty,deecke2018image,pidhorskyi2018generative}, but we consider AL, the more challenging task that is also more applicable to clinical practice.

The vast majority of AL methods benchmark on the industrial anomaly detection dataset \textit{MVTec-AD} \cite{bergmann2019mvtec}. Recent works include CutPaste, a self-supervised learning model which trains an encoder neural network to extract features that are useful for differentiating between normal and anomalous data on the proxy task of detecting the ``cutting and pasting'' of regions of images to another random part of the image \cite{li2021cutpaste}; PatchSVDD, a patch-based self-supervised model that utilizes support vector data descriptions (SVDDs) to detect and localize anomalies \cite{patchsvdd}; PatchCore, which localizes anomalies within patches by comparing their features to a memory bank of features of patches from normal images based on pretrained neural networks \cite{patchcore}; and PaDiM, a similar patch-based method that estimates the probability distribution of normal class instances \cite{defard2020padim}. 
These methods detect anomalies by comparing image features to features from normal training data; our method instead compares directly to a normal ``realization'' of the image given its surroundings, which is more robust to the high complexity of medical data, in particular breast tissue, where the possible feature similarity between anomalous and normal data can be much higher as compared to other types of data such as MVTec-AD. This is one possible reason for why these other methods, which perform very well on MVTec-AD, have a performance drop on medical data such as DBT, while our method performs significantly better.

Although MVTec-AD serves to model the application of anomaly localization to the industrial setting, the task of AL for medical images is also an important task for multiple reasons.
First, we find that AL methods that perform extremely well on MVTec-AD do \textit{not} necessarily translate well to medical image AL scenarios. This is due to the MVTec-AD data being significantly more controlled, less complex, and much less varied than the data seen in medical images, as well as having visual similarities to images from ImageNet, which have been absorbed by commonly-used pretrained image encoders, \ie , ResNet \cite{he2016resnet}. 
Healthy tissue in medical images often has high semantic variability, uncountably many distinct possible anomalies, and can generally be quite unpredictable, especially in highly variable anatomies like the breast. Simply put, many existing AL methods do not have the ability to fully generalize to the many possible challenging scenarios of medical anomaly detection. We believe that supporting a greater focus of general AD and AL research on the important, safety-critical application of medical imaging is essential for the development of methods that have broader impact.

\subsection*{Anomaly Localization with Image Completion}
Another direction of AL is to \textit{reconstruct} a test image and consider it as anomalous if the reconstruction is distinct from the input. Reconstruction-based methods, \eg, \cite{schlegl2019f,choi2019novelty,deecke2018image,pidhorskyi2018generative}, commonly solve this problem through an encoder-decoder mechanism.
These methods can not always be robust at discriminating anomalous data from normal data because given some input image, anomalous data within it may be partially reconstructed even by a normal-trained reconstructor, making anomalies not stand out within the reconstruction error.
Image completion-based methods, \eg, \cite{haselmann2018anomaly,munawar2015structural,pirnay2021inpainting,zavrtanik2021reconstruction,intra,swiecicki2021generative}, alleviate this problem by excluding the reconstructed region as input, and creating a normal completion that is more noticeably different than the anomalous ground truth.

Recent works \cite{wan2021high,liu2021pd,zhao2020uctgan,zheng2019pluralistic,dupont2019probabilistic} approached the goal of producing multiple plausible and diverse completions for a single input. 
As such, these methods are unnecessary and impractical for our purposes, which we show experimentally in Section \ref{sec:exp:compute}.

Instead, we achieve completion variability by a simple and novel application of spatial dropout layers to a pretrained completion network. Our method requires no additional training, and could theoretically be used on any sort of convolutional deterministic completion network. This keeps our overall anomaly detection method straightforward and intuitive, and importantly, fast.

\section{Methods}
\label{sec:methods}
\subsection{Introduction}
\label{sec:methods:intro}

\begin{figure}
\centering
\includegraphics[width=0.98\linewidth]{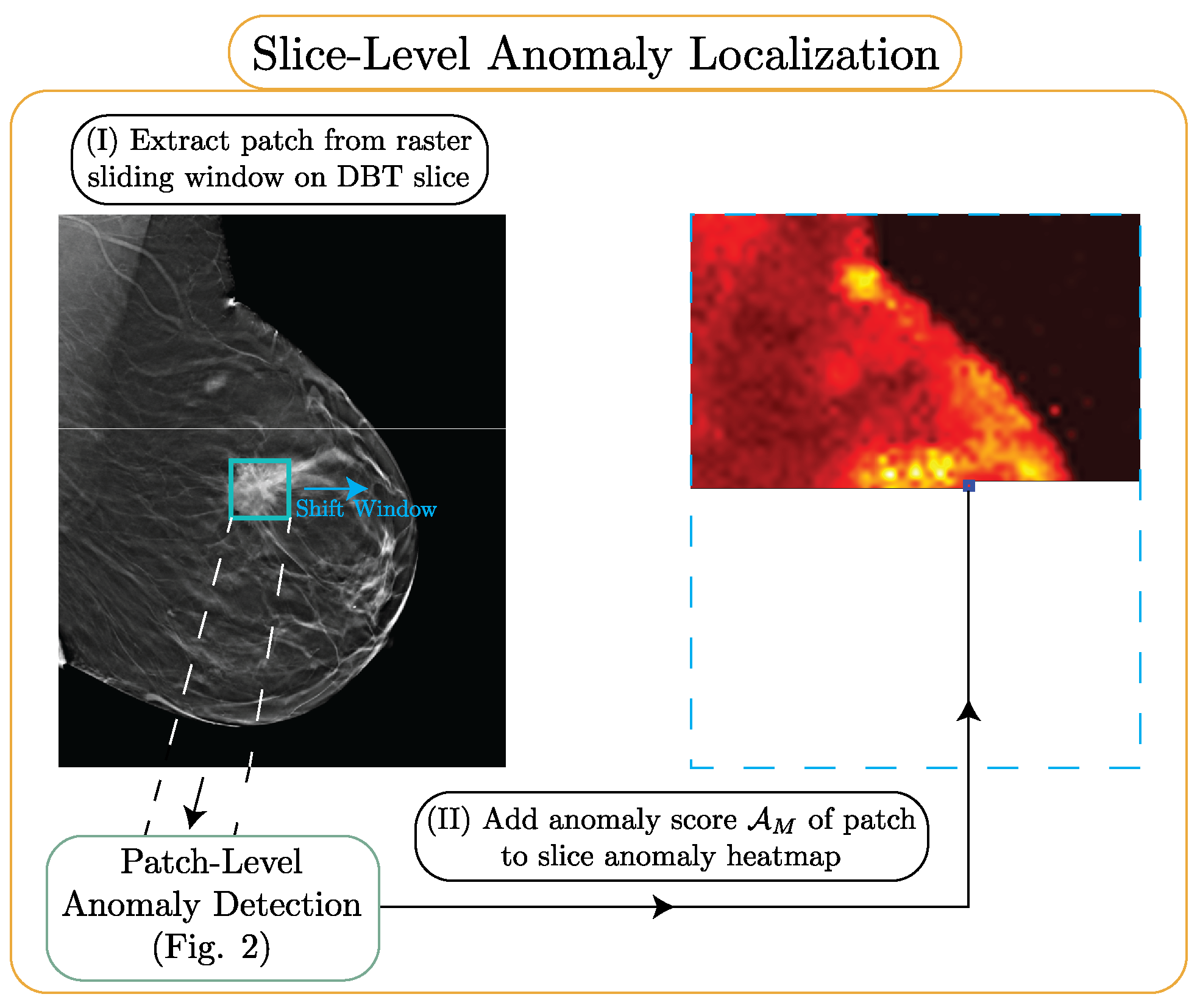}
\caption{\textbf{The outer loop of our proposed anomaly localization method}, PICARD (Algorithm \ref{alg:mindist}), at the \textit{slice} level. See Figure \ref{fig:model_patch} for the inner loop at the \textit{patch} level.}
\label{fig:model_slice}
\end{figure}

\begin{figure*}
\centering
\includegraphics[width=0.8\linewidth]{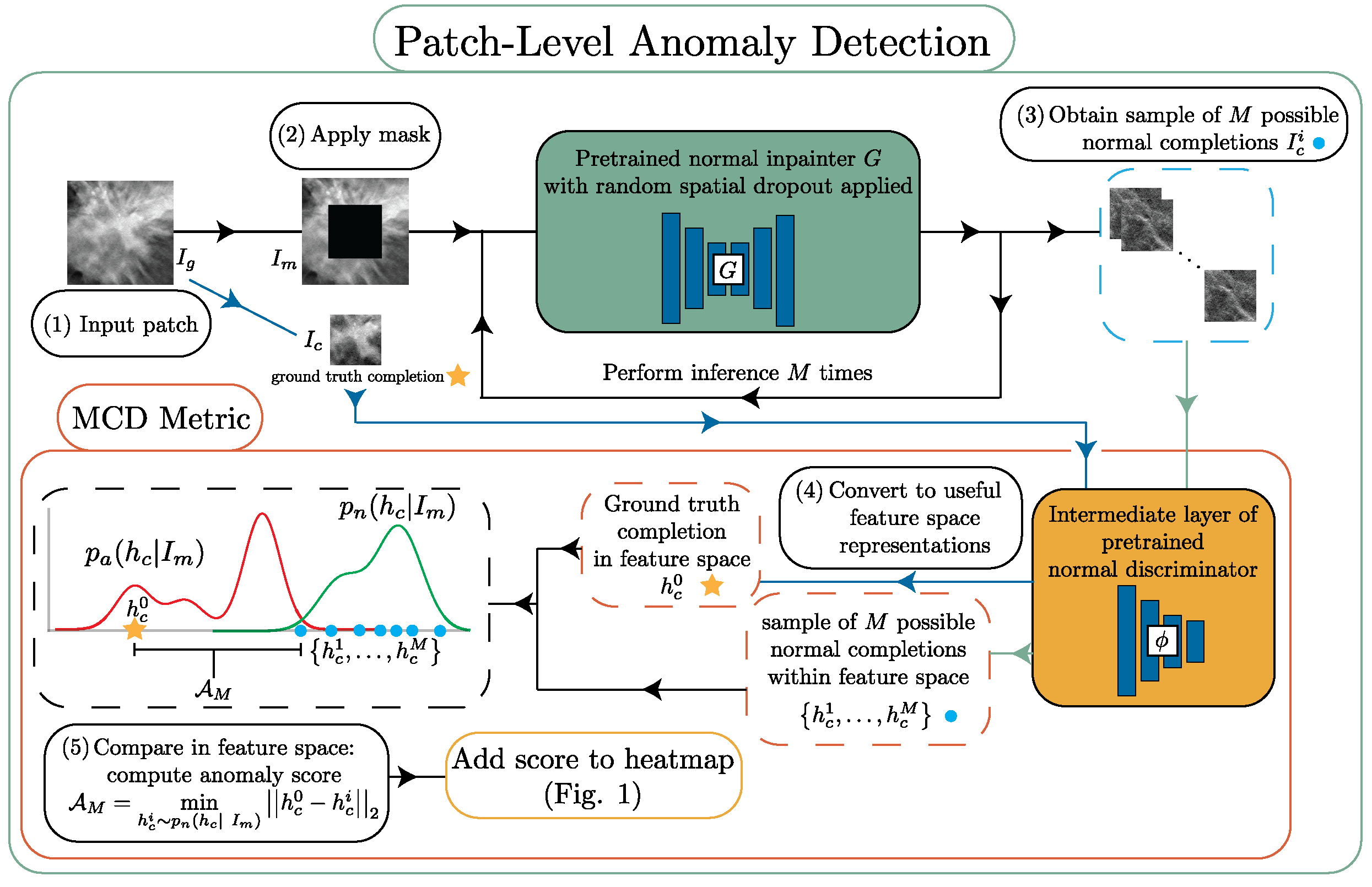}
\caption{\textbf{The inner loop of our proposed anomaly localization method}, at the \textit{patch} level. See Figure \ref{fig:model_slice} for the outer loop at the \textit{slice} level.}
\label{fig:model_patch}
\end{figure*}

Our overall anomaly localization method is summarized in Figures \ref{fig:model_slice} (outer, image/slice-level loop) and \ref{fig:model_patch} (inner, patch-level loop). Beginning with some 2D slice of a DBT scan volume, our model creates a ``sliding'' patch window that rasters through the slice with a fixed stride. For each sliding-window image patch $I$, we apply a mask over the center region $I_c\subseteq I$ to obtain an image $I_m= I - I_c$ with the region missing; \rev{we save the missing region $I_c$ as the \textit{ground truth completion}.} Next we compare the distribution of predicted normal completions of $I_m$ to the ground truth completion $I_c$, by examining the $L_2$ distance, in a feature space, of the prediction closest to the ground truth. If that distance is above a certain threshold, then $I_c$ is anomalous. In Figure \ref{fig:model_patch}, $p_a(h_c|I_m)$ and $p_n(h_c|I_m)$ are the feature space distributions of (1) anomalous and (2) normal completions of $I_m$, respectively. The anomaly score for each $I_c$ is used for the associated locations of the anomaly heatmap of the entire DBT slice (Fig. \ref{fig:model_slice}).

\subsection{The MCD Anomaly Metric and its Convergence Properties}
\label{sec:theory:auc_proof}

In this section we present the formal definition of the minimum completion distance (MCD) metric, which we use for anomaly detection at the patch level. Anomaly localization is then performed for an entire DBT slice by using the MCD metric on overlapping patches of that slice.

Consider a single test image/patch $I$ with a ground truth completion region $I_c$ that has surroundings $I_m$, \textit{i.e.}, $I=I_c\cup I_m$ and $I_c\cap I_m=\varnothing$. Now, consider sampling $M$ i.i.d. (independent and identically distributed) possible normal completions of $I_m$: $\{I_c^1,\ldots,I_c^M\}\sim p_n\left(I_c\middle|I_m\right)$. Here $p_n\left(I_c\middle|I_m\right)$ is the probability density function (p.d.f.) of the distribution of possible normal completions of $I_m$; we denote $p_a\left(I_c\middle|I_m\right)$ as the same but for \textit{anomalous} completions.

Now, use a pretrained normal data encoder $\phi$ to map the completions to a feature space via $h_c^i=\phi(I_c^i)$. 
Assuming $\phi$ to be an injective function, we can construct p.d.f.s of completions within this feature space; \ie, the feature space p.d.f. paired with $p_n(I_c|I_m)$ is $p_n(h_c|I_m)$. As such, $\phi$ transforms the completion image samples $\{I_c^i\}_{i=1}^M$ to $\{h_c^i\}_{i=1}^M\sim p_n\left(h_c\middle|I_m\right)$.

We then define the MCD anomaly score of $I$ to be
\begin{equation}
\label{eq:score}
\displaystyle\mathcal{A}_M\left(I_c;I_m\right)\triangleq\min\limits_{h_c^i\sim p_n\left(h_c\middle|I_m\right)}\left|\left|h_c^0-h_c^i\right|\right|_2,
\end{equation}
where $h_c^0=\phi\left(I_c\right)$ is the ground truth completion in feature space.

Next, we show that this metric is an arbitrarily powerful anomaly classifier as the sample size $M$ approaches $\infty$, and more practically that the performance improves with high sample efficiency, given reasonable assumptions about how the distributions of anomalous and normal data are distanced from each other. These assumptions are adapted from \textit{Proposition 2} of \cite{DUIAD}, and are summarized as follows:
\paragraph{Key Assumptions Appropriate for Anomaly Detection}
Given any test image $I$ with completion region $I_c$ and surroundings $I_m$, the feature-space distributions of possible normal and anomalous completions of $I_m$, $p_n(h_c|I_m)$ and $p_a(h_c|I_m)$, respectively, are sufficiently distant such that for most $h^0_c\sim p_n(h_c|I_m)$, $p_a(h^0_c|I_m)$ is small enough so that $p_a(h^0_c|I_m) \leq p_n(h^0_c|I_m)$ almost everywhere (see the dashed-line box at the bottom left of Fig. \ref{fig:model_patch}).

The AUROC/AUC, or Area Under the Receiver Operating Characteristic Curve, is a widely-used method for quantifying the performance of a classifier. One way of defining it is that the AUC is the probability of a positive sample being given a score higher than a negative sample \cite{fawcett2006ROC, yuan2020robust}. Consider some patch $I^n$ with completion region $I_c^n$ and surroundings $I_m^n$ that has no anomalies within $I_c^n$, and some other patch $I^a$ with completion region $I_c^a$ and surroundings $I_m^a$ that \textit{does} have anomalies within $I_c^a$. In this case, the definition of the AUC translates to the probability that the patch with an anomalous completion region will be scored higher than the patch with a normal completion region, i.e., $\mathrm{AUC} = \mathrm{Pr}\left(\mathcal{A}_M\left(I_c^a;I_m^a\right)>\mathcal{A}_M\left(I_c^n;I_m^n\right)\right)$. To optimize our anomaly metric for a given patch, we would like to maximize this equation. 
Next, we will evaluate the asymptotic performance of our novel metric's AUC with respect to $M$, in order to provide a formal analysis of our method's performance.  

\paragraph{Convergence Derivation} 
To begin, for readability we will define the minimum distance anomaly scores of $I^n$ and $I^a$ respectively as
$\epsilon_M^n=\mathcal{A}_M(I^n_c;I^n_m)$ and $\epsilon_M^a=\mathcal{A}_M(I^a_c;I^a_m)$ via Equation \eqref{eq:score}. 
Note that $\left\{h_c^1,\ldots,h_c^M\right\}\sim p_n\left(h_c\middle|I^n_m\right)$ are i.i.d. random variables. The normal ground truth $h_c^n=\phi(I_c^n)$ can also be thought of as being sampled from $p_n\left(h_c\middle|I^n_m\right)$ because it is just another valid completion of $I^n_m$;
similar reasoning applies to the anomalous ground truth $h_c^a=\phi(I_c^a)$ and $p_a\left(h_c\middle|I^a_m\right)$. As such, $\epsilon_M^n$ and $\epsilon_M^a$ are both continuous random variables (as both are functions of continuous random variables).
We then have 
\begin{align}
&\mathrm{Pr}\left(\epsilon_M^a>\epsilon_M^n\right)
=\int_{\epsilon_M^a=0}^{\infty}\int_{\epsilon_M^n=0}^{\epsilon_M^a}p\left(\epsilon_M^a,\epsilon_M^n\right)d\epsilon_M^nd\epsilon_M^a\\
&=\int_{0}^{\infty}{p\left(\epsilon_M^a\right)\int_{0}^{\epsilon_M^a}{p\left(\epsilon_M^n\right)d\epsilon_M^nd\epsilon_M^a}}, \label{eq:AUC_prob}
\end{align}
where the second line was obtained because $\epsilon_M^n$ and $\epsilon_M^a$ are independent, as they are respectively generated from possible completions of independent images. 

The inner integral $\int_{0}^{\epsilon_M^a}{p\left(\epsilon_M^n\right)d\epsilon_M^n}$ is the cumulative density function of $\epsilon_M^n$ evaluated at some given $\epsilon_M^a$,
\begin{align}
\int_{0}^{\epsilon_M^a}{p\left(\epsilon_M^n\right)d\epsilon_M^n}&=\mathrm{Pr}\left(\epsilon_M^n\le\epsilon_M^a|\epsilon_M^a\right)\\
&=1-\mathrm{Pr}\left(\epsilon_M^n>\epsilon_M^a|\epsilon_M^a\right). \label{eq:2}
\end{align}

Now, $\mathrm{Pr}\left(\epsilon_M^n>\epsilon_M^a|\epsilon_M^a\right)$ is the probability that out of the sample $\left\{h_c^1,\ldots,h_c^M\right\}\sim p_n\left(h_c\middle|I_m^n\right)$, there is no $h_c^i$  for $i=1,\ldots,M$ such that $\left|\left|h_c^n-h_c^i\right|\right|_2\le\epsilon_M^a$, \textit{i.e.} $\left|\left|h_c^n-h_c^i\right|\right|_2>\epsilon_M^a\forall i=1,\ldots,M$. This probability can therefore be computed as
\begin{align}
&\mathrm{Pr}\left(\epsilon_M^n>\epsilon_M^a|\epsilon_M^a\right)\\
&=\mathrm{Pr}\left(\left|\left|h_c^n-h_c^i\right|\right|_2>\epsilon_M^a,\forall i=1,\ldots,M\right)\\
&=\mathrm{Pr}\left(\left|\left|h_c^n-h_c^1\right|\right|_2>\epsilon_M^a\right)\times\cdots\\
&\qquad\cdots\times \mathrm{Pr}\left(\left|\left|h_c^n-h_c^M\right|\right|_2>\epsilon_M^a\right)\\
&=\prod_{i=1}^{M}\mathrm{Pr}\left(\left|\left|h_c^n-h_c^i\right|\right|_2>\epsilon_M^a\right)\\
&=\prod_{i=1}^{M}\left[1-\mathrm{Pr}\left(\left|\left|h_c^n-h_c^i\right|\right|_2\le\epsilon_M^a\right)\right], \label{eq:prod}
\end{align}
where the product expansion can be taken because each feature space completion sample $h_c^i$ of $I_m^n$ is independent. The term $\mathrm{Pr}\left(\left|\left|h_c^n-h_c^i\right|\right|_2\le\epsilon_M^a\right)$ within the product is the probability that the feature space distance between the (fixed) ground truth of the completion region and the (random) $i^{th}$ possible normal completion sample is less than the given $\epsilon_M^a$. This is found by integrating the probability density of normal completions (in feature space) that all of the $h_c^i$ were sampled from, $p_n\left(h_c\middle|I_m^n\right)$, over the ``$\epsilon$-ball’’ $B\left(h_c^n,\epsilon_M^a\right)$ with $\epsilon=\epsilon_M^a$ centered at $h_c^n$, defined by $B\left(h_c^0,\epsilon\right)=\left\{h_c:\left|\left|h_c^0-h_c\right|\right|_2\le\epsilon\right\}$.

This integral can be written as 
\begin{align}
\mathcal{P}(\epsilon_M^a)\triangleq\int_{B\left(h_c^n,\epsilon_M^a\right)}{p_n\left(h_c\middle|I_m^n\right)dh_c}, 
\end{align}
allowing Eq. \eqref{eq:prod} to become
\begin{align}
\label{eq:prod_ep}
\mathrm{Pr}\left(\epsilon_M^n>\epsilon_M^a|\epsilon_M^a\right)=\prod_{i=1}^{M}\left[1-\mathcal{P}(\epsilon_M^a)\right]=\left[1-\mathcal{P}(\epsilon_M^a)\right]^M,
\end{align}
as the integral is the same for all samples $h_c^i$ of possible normal completions of the masked image $I_m^n$, because it only depends on $h_c^n$ and $\epsilon_M^a$, the latter of which is computed with samples from the distribution of normal completions of the \textit{other} masked image $I_m^a$. 

As such, Equation \eqref{eq:2} can be written as
\begin{align}
\int_{0}^{\epsilon_M^a}{p\left(\epsilon_M^n\right)d\epsilon_M^n}=1-\left[1-\mathcal{P}(\epsilon_M^a)\right]^M,
\end{align}
which can be substituted into Eq. \eqref{eq:AUC_prob} to give
\begin{align}
&\mathrm{Pr}\left(\epsilon_M^a>\epsilon_M^n\right)=
\int_{0}^{\infty}{p\left(\epsilon_M^a\right)\left[1 - \left[1-\mathcal{P}(\epsilon_M^a)\right]^M\right]d\epsilon_M^a}\\
&=\int_{0}^{\infty}{p\left(\epsilon_M^a\right)d\epsilon_M^a}
-\int_{0}^{\infty} p\left(\epsilon_M^a\right)\left[1-\mathcal{P}(\epsilon_M^a)\right]^Md\epsilon_M^a\\
&=1-\int_{0}^{\infty} p\left(\epsilon_M^a\right)\left[1-\mathcal{P}(\epsilon_M^a)\right]^Md\epsilon_M^a,
\end{align}
written with an expectation value as
\begin{align}
&\mathrm{Pr}\left(\epsilon_M^a>\epsilon_M^n\right)
=\displaystyle 1-\mathop{\mathbb{E}}\limits_{\epsilon_M^a\sim p\left(\epsilon_M^a\right)}\left[1-\mathcal{P}(\epsilon_M^a)\right]^M.  \label{eq:auc_expec}
\end{align}

Recall that our goal is to evaluate the limit of the MCD metric AUC (Eq. \eqref{eq:auc_expec}) with respect to the normal completion sample size $M$.
Note that the normal completion probability density integral term within the expectation, $\mathcal{P}(\epsilon_M^a)$,
is bounded by $(0,1)$ for all $M$ because $\epsilon^a_{0,M}$ is bounded by $(0,\infty)$, as $\epsilon_M^a\neq 0$ is a zero probability event where a sampled $h^j_c$ is exactly $h^a_c$. This means that $1-\mathcal{P}(\epsilon_M^a)$ is also bounded by $(0,1)$, and therefore so is $[1-\mathcal{P}(\epsilon_M^a)]^M$, which means that the expectation of Eq. \eqref{eq:auc_expec} is as well. As $\lim\limits_{M\rightarrow\infty} \alpha^M = 0$ for all $\alpha\in (0,1)$, then it must be the case that $\lim\limits_{M\rightarrow\infty}\mathbb{E}_{\epsilon_M^a\sim p\left(\epsilon_M^a\right)}[1-\mathcal{P}(\epsilon_M^a)]^M=0$, so that
\begin{align}
&\lim\limits_{M\rightarrow\infty}\mathrm{Pr}\left(\epsilon_M^a>\epsilon_M^n\right)=\\
&=\displaystyle 1-\lim\limits_{M\rightarrow\infty}[1-\mathcal{P}(\epsilon_M^a)]^M = 1 - 0 = 1,% \label{eq:qed}
\end{align}
\textit{i.e.} the classifier theoretically approaches a perfect AUC as the sample size $M\rightarrow\infty$. We evaluate this behavior empirically in Section \ref{sec:exp:vsM}.

However, for practical purposes we must consider how this score performs for a reasonably-sized $M$; from Eq. \eqref{eq:auc_expec} we need $[1-\mathcal{P}(\epsilon_M^a)]^M$ to be sufficiently close to zero for a low enough $M$. % as $M\rightarrow\infty$. 

We can achieve a better empirical performance guarantee
by applying the previously stated assumptions of anomaly detection. First note that as $M$ increases, $[1-\mathcal{P}(\epsilon_M^a)]^M$ will continually decrease, as $\epsilon_M^a$ remains the same or decreases as $M\rightarrow\infty$. A useful AUC for a satisfactorily-low $M$ will occur if the term $[1-\mathcal{P}(\epsilon_M^a)]^M$ decreases quickly as $M$ increases. In fact, our earlier work of \cite{swiecicki2021generative} is built on the case of $M=1$. Through our spatial dropout method that we use for pluralistic completions (Section \ref{sec:theory:dropout_inpaint}), our method can generate any $M$ unique completions, thus allowing for a higher AUC.

By assumption, the normal and anomalous distributions $p_n(h_c|I_m^a)$ and $p_a(h_c|I_m^a)$, respectively, are sufficiently distant that it is unlikely that samples from one will be close to samples from the other. Now, consider steadily incrementing $M$ from $1$. For an anomalous $h^a_c$, $\epsilon_M^a$ will begin large, and although likely getting slightly smaller as $M$ increases due to additional samples, it is expected to stay reasonably large, making $[1-\mathcal{P}(\epsilon_M^a)]^M$ decrease quickly. Even if some sample $h^i_c\sim p_n(h_c|I_m^a)$ happens to be close to $h^a_c$, any sampling where this is non-trivially likely to happen will have high enough $M$ for $[1-\mathcal{P}(\epsilon_M^a)]^M$ to already be very small despite the accompanying low $\epsilon_M^a$, so this is a non-issue. It is also very unlikely for $\epsilon_M^a$ to begin small, although this would be a ``failure mode'' as $\epsilon_M^a$ would only decrease slowly from there. In summary, we should achieve good performance for a low $M$; indeed we find in practice that $M=10$ is sufficient to achieve beyond state-of-the-art anomaly localization results, and the influence of different choices of $M$ is shown in Section \ref{sec:exp:vsM} as well.

\subsection{Completion Variability with Spatial Dropout}
\label{sec:theory:dropout_inpaint}

We have shown theoretical support for our MCD anomaly metric, and explained why given appropriate assumptions of the distributions of normal and anomalous data, the metric performs well for a reasonably small completion sample size $M$. The central component of this metric is the diverse sampling from $p_n(I_c|I_m)$: the distribution of possible normal completions $I_c$ of some surroundings $I_m$; in other words, obtaining \textit{pluralistic completions}. 
We opt for a simple and intuitive approach for creating pluralistic completions that still manages to achieve sufficient feature variability of completions for the MCD metric. 
Our goal is to make the output of a completion network $G$ trained on normal data \textit{variable} for some fixed input masked image $I_m$. At each $i^{th}$ evaluation of $G(I_m)$ we wish to obtain a different output completion $I^i_c$, while still maintaining the ability of $G$ to create fairly realistic normal inpaintings that will be distinct from any anomalous data.

Our intuition is from the dropout \cite{srivastava2014dropout} mechanism, which is commonly used during training to combat overfitting, but can also be used during inference to produce variable outputs \cite{kendall2017uncertainties}. 
We apply this general prescription to a completion network to induce variability for conditional generative models.
This idea is briefly introduced in \cite{wieluch2019dropout}, but they only present it as a proof-of-concept extension of their work, while we manage to implement it in a real application setting.

In particular, we perform completions using the model of \cite{yu2018generative}, which includes a \rev{Wasserstein generative adversarial network (GAN)-based} fully-convolutional completion network $G$ and critic/encoder $\phi_W$ \cite{arjovsky2017wasserstein}.
$G$ creates a fixed completion $I_c$ of some input masked image $I_m$, while $\phi_W$ learns to discriminate between real vs. fake normal completion data; we use $\phi_W$ as the completion feature encoder $\phi$ described in Section \ref{sec:theory:auc_proof}. \rev{Wasserstein generative adversarial networks (GANs) are signicantly more reliable to train than traditional GANs \cite{goodfellow2014generative}: they converge reliably, remove problems such as mode collapse, and have interpretable loss functions, among other benefits. The critic $\phi_W$ is the Wasserstein GAN's version of the traditional GAN's discriminator.}

\begin{figure*}
    \centering
    \includegraphics[width=0.7\linewidth]{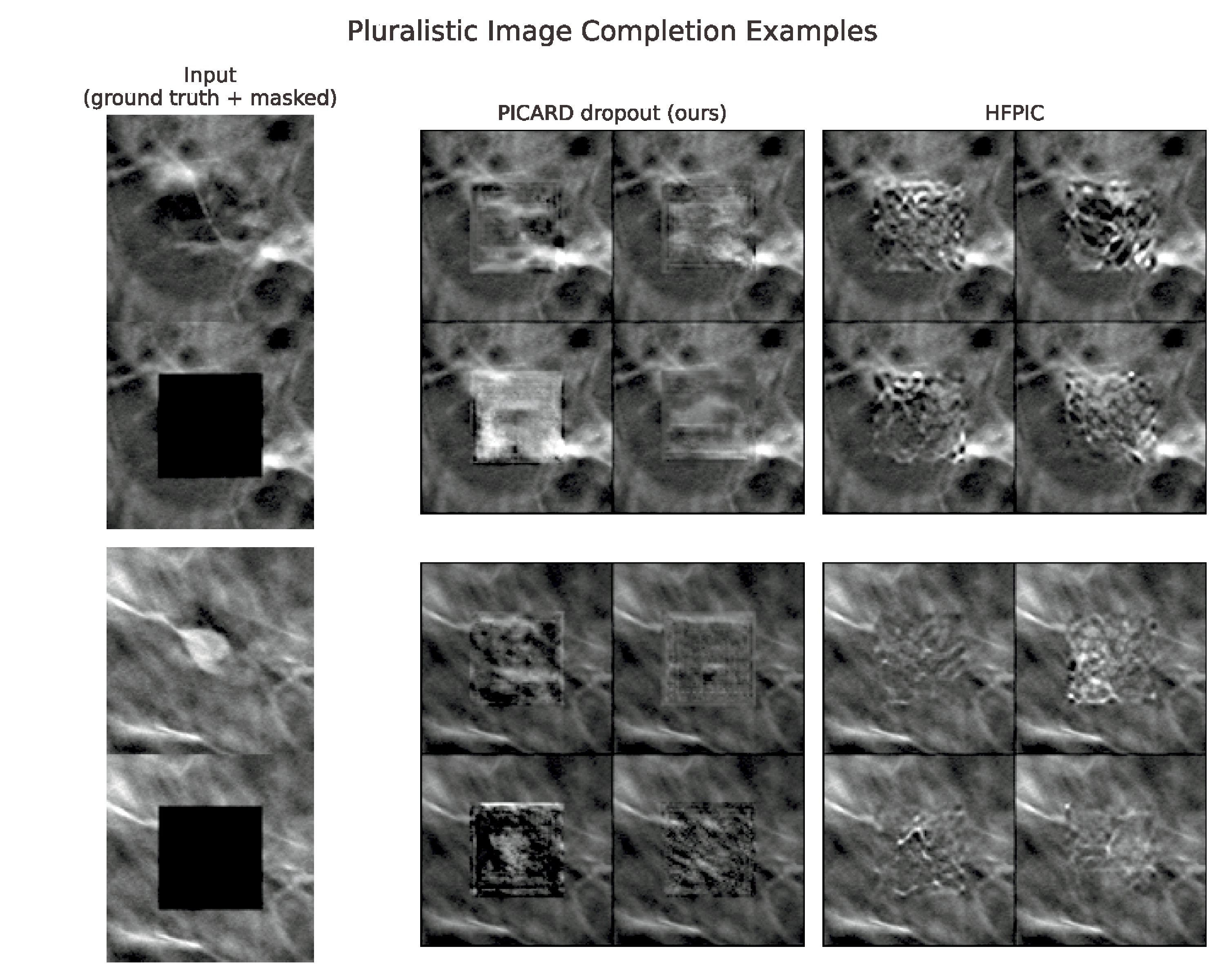}
    \caption{\textbf{Examples of pluralistic normal completions} of a normal DBT patch (top block) and an anomalous patch (bottom block) using our method (Section \ref{sec:theory:dropout_inpaint}) and HFPIC \cite{wan2021high}. Left column: input image, masked and unmasked; center column: completions with our method; right column: completions with HFPIC. \revfinal{Image contrast modified to improve visibility.}}

    \label{fig:eg_completions}
\end{figure*}

In our setting, since the input to $G$ is $I_m$, a fixed variable, the network has no inherent stochasticity by default. A simple method of adding stochasticity to the input, \ie, $G(I_m, z)$ where $z$ is sampled from some noise distribution, would not work as well because the network would simply learn to ignore $z$ \cite{isola2017image, mathieu2015deep}.
In order to allow $G$ to create semantically diverse yet sufficiently high-quality completions at each evaluation of a single $I_m$, we propose using \textit{spatial}, or \textit{channel-wise} dropout \cite{spatialdropout} within $G$. This type of dropout randomly makes entire channels of convolutional layer activation maps zero with some probability, rather than individual neurons. We give examples of pluralistic completions using our method in Figure \ref{fig:eg_completions}. \rev{We note that using dropout on $G$ naturally results in reduced visual quality of individual completions, due to the variability that dropout adds to the network. However, we found that the benefit of having access to multiple possible completions outweighs this, still resulting in improved tumor detection performance over the single-completion case (Table \ref{tab:modelcompare}).}

Conceptually, because convolutional layer activation maps carry spatial correlations between adjacent pixels, \rev{dropping out individual activations randomly can result in low-quality completions, which we found to be the case in practice}. On the other hand, dropping an entire channel of an activation map with spatial dropout, can be thought of as inducing a change in the global feature information of the resulting completion, while avoiding any such negative spatial effects. This makes intuitive sense at a high level: for a given layer of a \textit{fully connected} neural network, the individual neuron's activations are the key global features that affect the downstream inference; on the other hand, for a convolutional neural network layer, the key global features are different channels of the given activation map. \cite{lee2020revisitingspatialdrop} in fact found that spatial dropout used on convolutional neural networks (CNNs) can be functionally similar to using regular dropout on fully connected neural networks.

Intriguingly, we did not find any benefit in the quality or diversity of pluralistic completions between the options of (1) using dropout on $G$ during both training and testing or (2) only testing, over the course of many experiments.
As such, for the sake of simplicity, we obtain pluralistic completions by only applying dropout at test time to a normally (non-dropout) trained completion network. We also obtained better inpainting quality (on the training set) when dropout is excluded from the shallowest and deepest layers of $G$ (see Appendix A.1 for details). For all experiments we use a dropout probability of $0.5$, \rev{a relatively high value that we found suitable for generating sufficiently diverse completions, which was also assisted by dropout being used after the majority of $G$'s layers. We found that different training iterations of $G$ sometimes resulted in differing quality and variability of completions once dropout was applied, but we saw no obvious trends to this, so chose to halt training simply when the $L_1$ distance between completions and ground truth was minimized (see Section \ref{sec:exp:comparison} for more training details).}

\rev{It is also conceivable to explicitly optimize the placement and probabilities of dropout layers to maximize completion diversity and visual quality. However, there is not an explicit, differentiable dependence of a completion diversity metric (\eg, LPIPS \cite{zhang2018lpips}) or quality metric (\eg, the critic/discriminator score) on the dropout layer parameter(s) and/or placement, so it is unclear how these parameters could be efficiently tuned to optimize for these metrics. Doing so would require a non-differentiable optimization method such as Bayesian optimization, which is computationally prohibitive due to the high number of possible dropout parameters to tune (layer-by-layer), and the computational cost of sampling enough completions at each iteration of such a routine to get a reasonable holistic measure of completion diversity and quality. We attempted this Bayesian optimization procedure in early experiments but found that it did not converge; due to these issues, we simply fixed the dropout probability to a fixed value (0.5) for all layers, which we found sufficient for completion diversity and quality.}

\rev{Our} method also has the added \rev{benefit} that only negligible additional computational load is needed to create a pluralistic completion compared to an \rev{ordinary} deterministic completion network, as each \rev{completion} is created by an independent forward pass through $G$. \rev{This also makes pluralistic completion sampling easily parallelizable.}

\subsection{Full Anomaly Localization Method: PICARD}

We can now determine whether some $d_p\times d_p$ patch of a DBT scan includes an anomaly within the center square $d_m\times d_m$ region by sampling $M$ possible normal completions of that region given the surroundings, and using the minimum completion distance (MCD) metric (Eq. \eqref{eq:score}) to compare the completions to the missing region ground truth. The final portion of our model is to use this new metric to \textit{localize}, or segment, anomalies within a full size DBT slice.

Anomaly localization requires synthesizing an anomaly \textit{heatmap} for a given DBT slice $X$ that is the same size as that slice, where each pixel of the heatmap corresponds to the model's prediction confidence of the corresponding slice pixel containing anomalous data. To do so, we begin with the $d_p\times d_p$ image patch at the top left of $X$---our ``window''---and apply the MCD metric to that patch, with the aforementioned masked region chosen \textit{a priori}, to obtain an anomaly score associated with that patch. We then shift the window by some stride according to a basic overlapping raster scan order, perform the same procedure to obtain an anomaly score for this next window, and repeat until all raster windows have been scored, ending with the patch at the bottom right of $X$. The heatmap is all of these scores arranged with the same spatial orientation of the corresponding raster patch centers that created the scores. Finally, we use bicubic interpolation to upsample the heatmap until it is of the same size as $X$. The overall anomaly localization procedure is summarized in Algorithm \ref{alg:mindist}. In practice, the two \textbf{for} loops are parallelized to take full advantage of GPU memory; \ie, multiple completions are sampled, for multiple inputs, all at once. This feature creates a large decrease in computation time compared to our previous work of \cite{swiecicki2021generative}, where completions were simply made one-at-a-time. 

\begin{algorithm}[tb]
    \caption{Pseudocode for PICARD MCD Anomaly Detection for DBT scans}
    \label{alg:mindist}
    \textbf{Input}: Input DBT scan $X$, patch size $d_p=256$, mask size $d_m=128$, pluralistic completion sample size $M=10$, and completion network $G$ with completion encoder/critic $\phi$, both pretrained on normal data. \\
    \begin{algorithmic}[1] %[1] enables line numbers
    \STATE Initialize raster scan order of ``sliding'' windows of size $d_p\times d_p$ and stride $32$, starting at the top left of $X$.
    \FOR{each window in raster scan order}
        \STATE Let $I$ be image within sliding window. Remove centered $d_m\times d_m$ mask from $I$ to obtain $I_c\subseteq I$, with remaining surroundings $I_m = I - I_c$.
        \FOR{$i = 1, \ldots , M$}
            \STATE \textit{Sample normal completion $I_c^i\sim p(I_c|I_m)$ with dropout probability $p_{drop}$ on $G$:}
            \STATE $I_c^i= G(I_m)$    
        \ENDFOR
        \STATE \textit{Convert ground truth $I_c$ and predictions $\{I_c^i\}_{i=1}^M$ to feature space via $\phi$:}
        \STATE $h^0_c=\phi(I_c), \quad h_c^i = \phi(I_c^i) \,\, \forall i = 1, \ldots , M$ 
        \STATE \textit{Compute MCD anomaly score for $I_c$ (Equation \eqref{eq:score}):}
        \STATE $\mathcal{A}_M=\min\limits_{i = 1, \ldots , M}\left|\left|h_c^0-h_c^i\right|\right|_2$
        \STATE Store $\mathcal{A}_M$
    \ENDFOR
    \STATE Use anomaly scores $\mathcal{A}_M$ for each raster window to create anomaly heatmap, maintaining 2D spatial orientation.
    \STATE Use bicubic interpolation to upsample heatmap to size of $X$.
    \RETURN anomaly heatmap for $X$
    \end{algorithmic}
\end{algorithm}

\section{Dataset}
\label{sec:data}

For all experiments we use full size 2D slices of breast cancer DBT Digital Breast Tomosynthesis (DBT) scans from the Breast Cancer Screening (BCS)-DBT dataset \cite{buda2021data}. The scans have resolutions of either $1,890\times 2,457$ or $1,996\times2,457$ pixels. For training all models we used $6,245$ healthy slices of DBT volumes from the training set of BCS-DBT, each of which come from a different anatomical view and/or patient. $256\times 256$-shaped patches are randomly sampled from these slices for pretraining the completion network $G$ and the encoder $\phi$ for PICARD. For testing we use $133$ DBT slices that each contain at least one radiologist-annotated tumor, obtained from the test set of BCS-DBT. \rev{The tumor bounding-box annotations in the test set range in size from about $0.2\%$ to $7\%$ of the total area of a DBT image.}  All DBT slices are left-aligned for symmetry. We provide more details for the creation of this dataset and the impact of using it to test anomaly localization in Appendix C. Code for reproducing all experiments will be made publicly available.

\begin{table}
    \centering
    \caption{Summary of the DBT data used in this work, from the BCS-DBT dataset \cite{buda2021data}.}
    \begin{tabular}{cc}
        \toprule
        Dataset & No. of DBT scan slices\\
        \midrule
        Training (healthy only) & 6,245 \\
        Testing (biopsied cancer only) & 133 \\
        \bottomrule
    \end{tabular}
    \label{tab:data}
\end{table}

\section{Experiments and Results}
\label{sec:experiments}
Given that the outputs from our method are pixel-wise heatmaps, and only ground truth lesion \textit{bounding boxes} are provided in BCS-DBT, we adopt pixel AUC/AUROC as our anomaly localization (AL) evaluation metric, as in other AL works \cite{schlegl2019f, patchsvdd, patchcore, li2021cutpaste, defard2020padim}. \rev{We note that such AL algorithms cannot be evaluated with object-level detection metrics such as IoU because they do not output binary localization predictions such as segmentations or bounding boxes, as tuning some method used to seperate the heatmap into foreground and background pixels and then form object boxes/masks would require a validation set containing labeled anomalies, which is not permissible within the AL/AD setting.}

\begin{figure}
\centering
\subfigure{\label{fig:disteg}\includegraphics[width=0.49\linewidth]{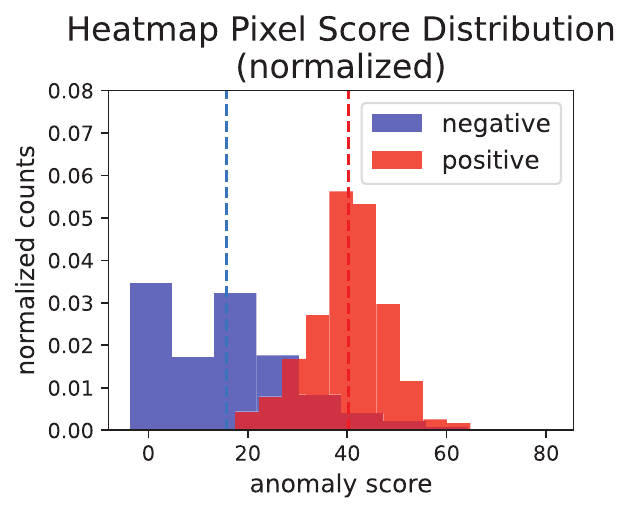}}
\subfigure{\label{fig:ROCeg}\includegraphics[width=0.49\linewidth]{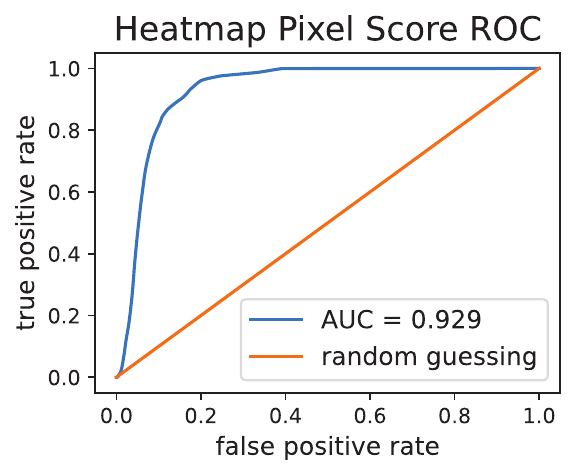}}
\caption{Histogram (left) and associated AUC (area under the receiver operating characteristic curve, right) of a particular test DBT cancer slice (the top left row of Fig. \ref{fig:comparemodels}), for the normalized distributions of MCD anomaly metric scores for normal pixels (blue) and anomalous pixels (red).}
\label{fig:pixelAUCeg}
\end{figure}

Specifically, we label all pixels of the slice as negative, except for the pixels inside and along the bounding box(es), which we label as positive. \rev{To obtain an anomaly localization/pixel AUC score for a given image, each pixel's binary label (normal or anomalous) is compared to the corresponding anomaly score from our model's predicted heatmap for that pixel, and the pixel is classified as anomalous if its score is above a certain threshold. The pixel-wise AUC for the image analyzes all possible score thresholds for a given image/heatmap to provide a holistic measure of anomaly localization performance for the entire image.} The final performance metric for the entire test set is the average pixel AUC of all slices. We also include a specific example of the associated ROC curve with the anomaly score distributions of normal and anomalous pixels created by using PICARD to heatmap a DBT slice from the test set in Figure \ref{fig:pixelAUCeg}.

\subsection{DBT Tumor Localization}
\label{sec:exp:comparison}
\begin{figure*}
\centering
\subfigure{\label{fig:qual0}\includegraphics[width=0.49\linewidth]{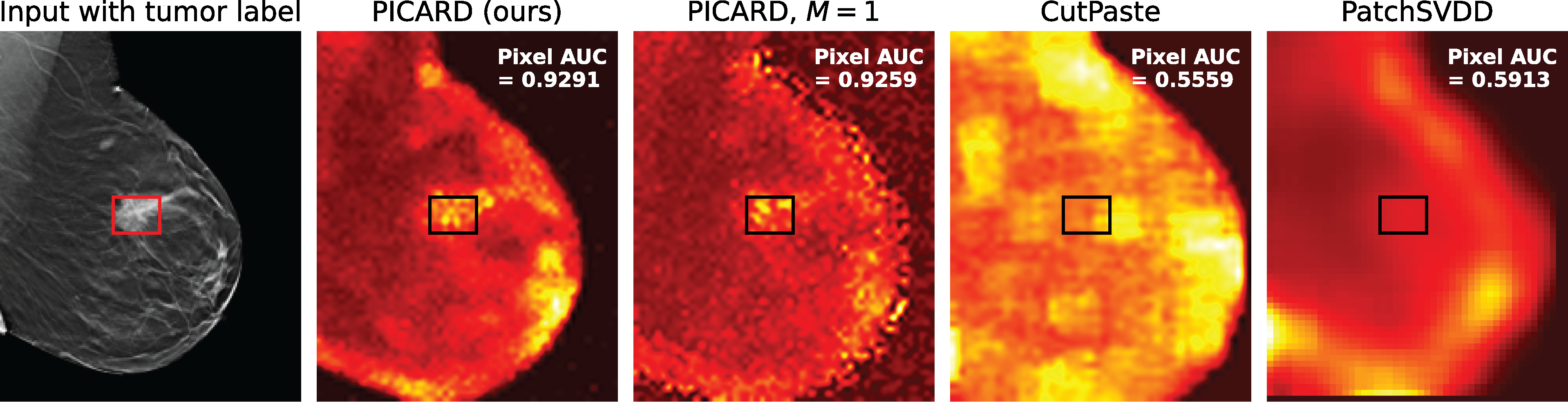}}
\subfigure{\label{fig:qual3}\includegraphics[width=0.49\linewidth]{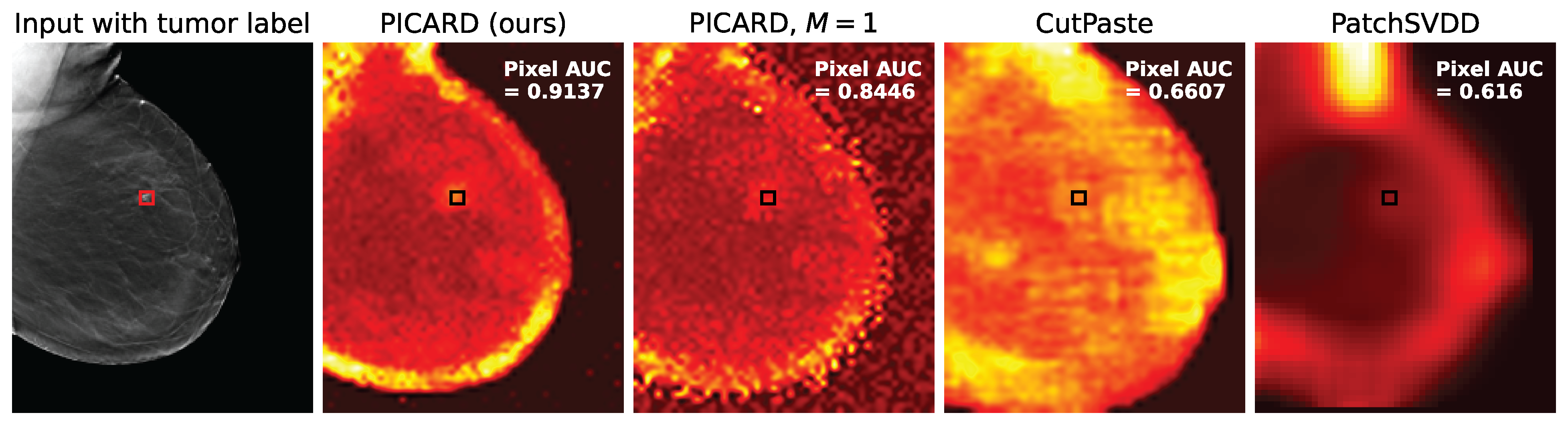}}
\subfigure{\label{fig:qual1}\includegraphics[width=0.49\linewidth]{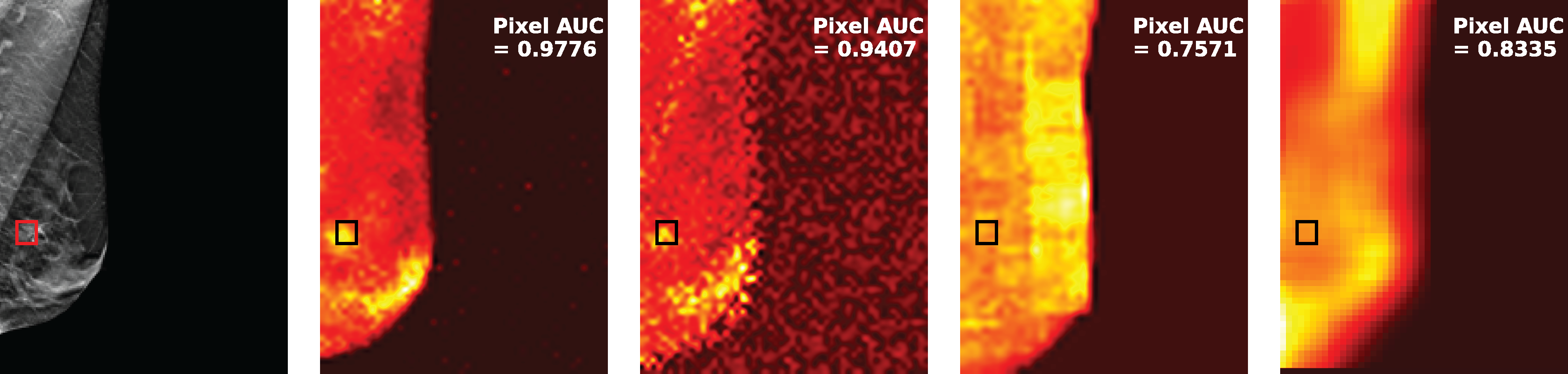}}
\subfigure{\label{fig:qual4}\includegraphics[width=0.49\linewidth]{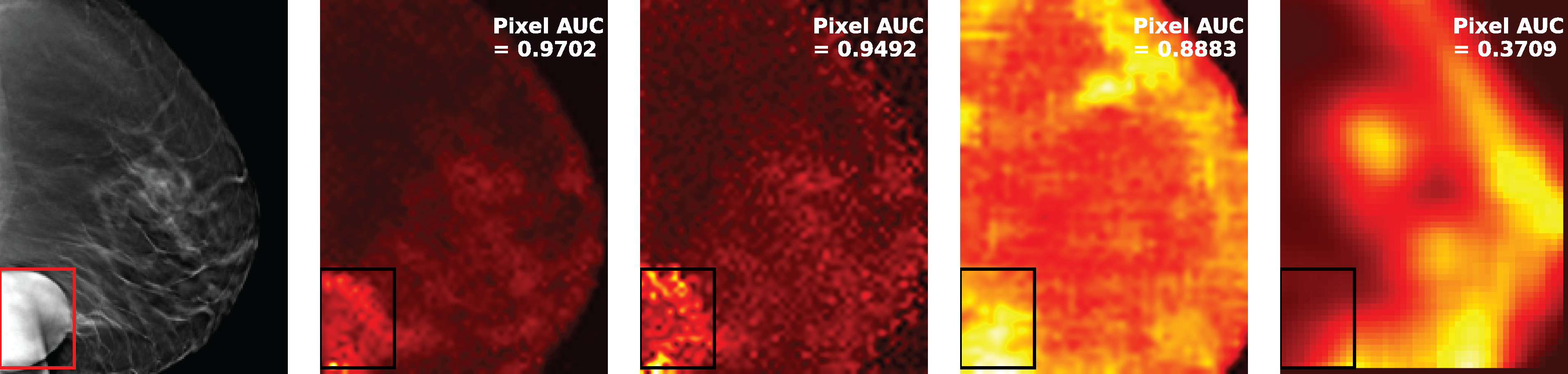}}
\subfigure{\label{fig:qual2}\includegraphics[width=0.49\linewidth]{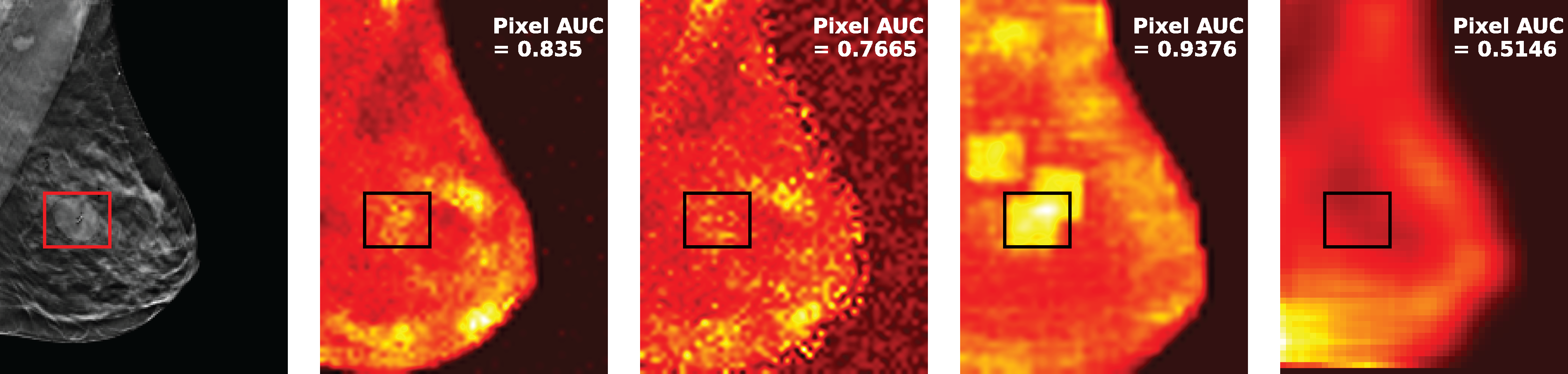}}
\subfigure{\label{fig:qual5}\includegraphics[width=0.49\linewidth]{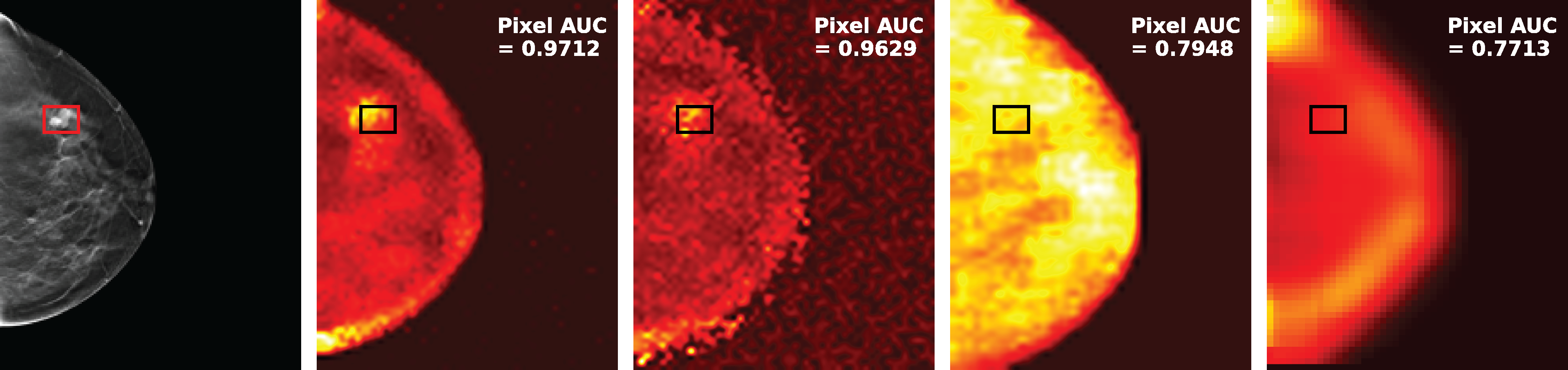}}
\subfigure{\label{fig:qual_dense1}\includegraphics[width=0.49\linewidth]{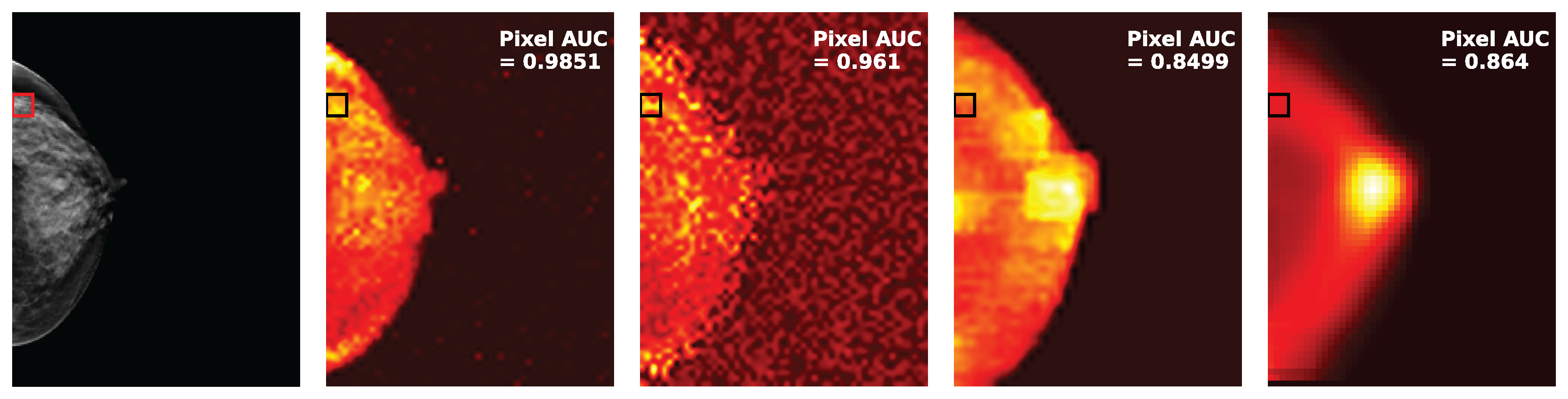}}
\subfigure{\label{fig:qual_dense2}\includegraphics[width=0.49\linewidth]{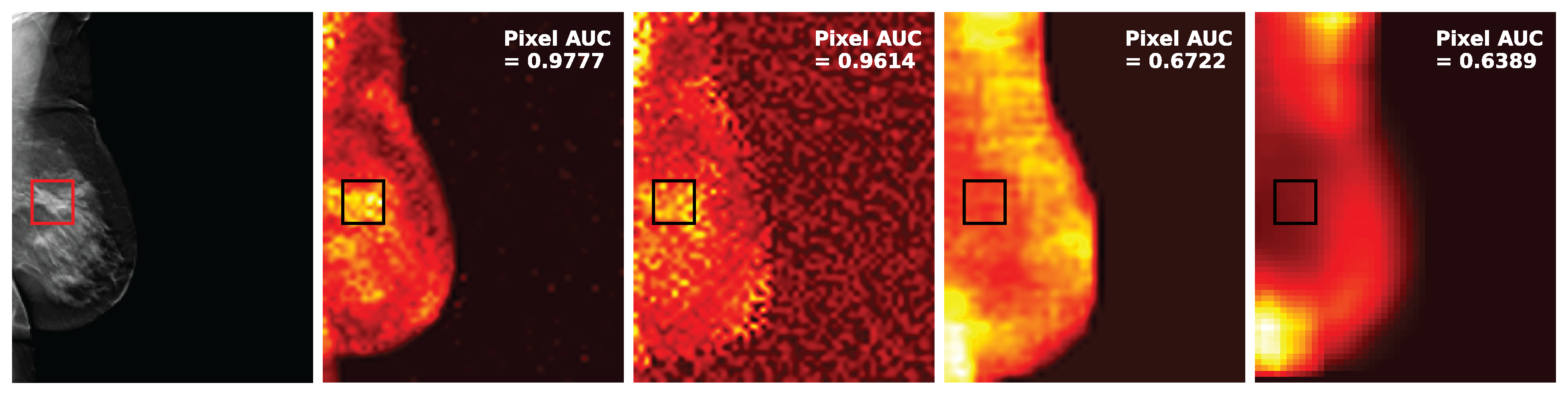}}
\caption{\textbf{Qualitative tumor localization performance for our method (PICARD) compared to several state-of-the-art methods.} \rev{For each example test image, we show the performance (from left to right) of} (1) our method, PICARD; (2) PICARD with the deterministic, single-completion case; (3) CutPaste \cite{li2021cutpaste}; and (4) PatchSVDD \cite{patchsvdd}. \rev{The two examples on the bottom row demonstrate performance on cases with dense breast tissue.} Refer to Table \ref{tab:modelcompare} for corresponding quantitative results on the entire test set. This figure is best viewed in color.}
\label{fig:comparemodels}
\end{figure*}

Now we compare previous leading unsupervised/self-supervised anomaly localization methods to our work: quantitative (pixel AUC) results are summarized in Table \ref{tab:modelcompare}, while qualitative (anomaly heatmap) results are given in Figure \ref{fig:comparemodels}. \rev{We also provide the pixel-wise \textit{average precision} (AP) score for each method in Table \ref{tab:modelcompare}. The AP summarizes the precision-recall curve, and is the weighted mean of the precision achieved at each possible scoring threshold along the precision-recall curve, according to scikit-learn's \texttt{sklearn.metrics.average\_precision\_score} in Python.} Further details and results of these comparison studies are given as follows, where we first explore other state-of-the-art methods, followed by our model.

\begin{table*}
    \centering
    \caption{Quantitative comparison of tumor localization methods on the DBT test set of cancerous scans.}
    \begin{tabular}{lccc}
        \toprule
        Method & Pixel AUC & Pixel AP & \begin{tabular}{@{}c@{}}Inference time \\ (per patch) (sec.)\end{tabular}\\
        \midrule
        \begin{tabular}{@{}l@{}}PICARD (ours)  \\ \textit{(image space)}\end{tabular} & \textbf{0.875} & \textbf{0.0943} & \textbf{0.062}\\
        \begin{tabular}{@{}l@{}}PICARD \\ \textit{(feature space)}\end{tabular}  & 0.865 & 0.0672 & 0.064 \\
        \begin{tabular}{@{}l@{}}PICARD, $M=1$  \\ \textit{(image space)}\end{tabular} & 0.846 & 0.0817 & \textbf{0.062} \\
        \begin{tabular}{@{}l@{}}PICARD, $M=1$ \\ \textit{(feature space)}\end{tabular} & 0.826 & 0.0582 & 0.064 \\ 
        PatchSVDD \cite{patchsvdd} & 0.777 & 0.0303 & 4.13 \\
        CutPaste \cite{li2021cutpaste} & 0.737 & 0.0522 & 0.087 \\
        \bottomrule
    \end{tabular}
    \label{tab:modelcompare}
\end{table*}

\paragraph{CutPaste \cite{li2021cutpaste}}
CutPaste first learns self-supervised deep representations and then builds a generative one-class classifier on learned representations. The representations are learned by classifying normal data from a novel data augmentation strategy that cuts an image patch and pastes it at a random location of a large image. To localize defective regions, CutPaste crops the images before applying the augmentation. CutPaste obtained leading anomaly localization results on the most common benchmark of MVTec-AD \cite{bergmann2019mvtec}.

To make a fair comparison to our method, we adopt the best performing augmentation strategy (\textit{CutPaste 3-Way}) and use the sliding-window hyperparameters as for PICARD. We further adjust the method to not select blank image patches or paste them in blank regions. These changes increase the classification difficulty and improve the performance. We train CutPaste until loss convergence, at $6,000$ epochs.

During inference, we extract embeddings from all patches with a given stride and learn a generative classifier at each location via a simple parametric Gaussian density estimator (GDE), with a log-probability density of
$\log p_{gde}(x_{ij}) \propto \{ -\frac{1}{2}(f(x_{ij})-\mu_{ij})^T\Sigma^T(f(x_{ij})-\mu_{ij}) \}$,
where $i,j$ specifics the spatial location for the current image patch $x_{ij}$, and $f$ is the feature embedding network. The final anomaly score map is obtained by accumulating prediction scores from all the generative classifiers. We find that CutPaste obtains an anomaly localization result on the DBT test set of $0.737$ average pixel AUC. A few example heatmaps are shown in Figure \ref{fig:comparemodels}.

\paragraph{PatchSVDD \cite{patchsvdd}}
PatchSVDD is an extension of DeepSVDD \cite{deepsvdd} to solve the problem of high-level intra-class variations by adopting patches, instead of entire images, as network inputs. It alleviates the collapse issue of mapping features to a single center by minimizing the distances between features extracted from spatially adjacent patches. The method also proposes an additional self-supervised learning task to predict the relative location of two nearby patches. 

Since this method also makes the predictions at patch level, we can directly adopt this method in our setting. To make it compatible with the DBT dataset, we select the same patch and stride size as PICARD, $256$ and $32$, respectively. During inference, we also use the same protocol PatchSVDD proposed to generate the anomaly map for every DBT image. We follow the same training parameters and procedure as in the original paper, and set the loss scaling hyperparameter $\lambda=1$. On the DBT test set, PatchSVDD achieves $0.777$ pixel AUC. Several example heatmaps are presented in Figure \ref{fig:comparemodels}.

\paragraph{PICARD (ours)}
Lastly, we evaluate our method, PICARD, (Figs. \ref{fig:model_slice}, \ref{fig:model_patch} and Algorithm \ref{alg:mindist}), on the DBT test dataset. 
We set patch size $d_p=256$ and mask size $d_m=128$. Empirically, this setting is sufficient enough to allow room for high resolution, variable completions to capture a variety of anomalies, while small enough to be able to precisely localization anomalies on the much larger, global DBT slices, which have resolutions of approximately $2,000\times 2,500$. For all experiments we set the raster window stride to $32$ pixels (so that the raster windows overlap), and we set completion sample size $M=10$. \rev{We trained the inpainter $G$ and critic $\phi$ with a batch size of $55$, halting once the $L_1$ training set reconstruction error between completions and their corresponding ground-truth images stopped decreasing.}  All experiments were completed on four 24 GB NVIDIA RTX 3090 GPUs. \rev{Each heatmap} took approximately two minutes to create, by processing multiple sliding raster window inputs, each sampling multiple completions, all in parallel. All other experimental and model training details are given in Appendix A.1.

On the test set, PICARD achieves an average pixel-level AUC for lesion detection of $\mathbf{0.875}$ with the MCD metric in image space, and $\mathbf{0.865}$ in feature space, outperforming other existing methods by at least $10\%$ AUC. We find that when we set $M=1$ in order to evaluate the single completion case, these values shift to $0.846$ and $0.826$ for image and feature space, respectively. These are the first BCS-DBT pixel AUC results for this method that was first introduced in our work of \cite{swiecicki2021generative}, which itself already beats other state-of-the-art methods. \rev{We also find that our approach similarly outperforms all other methods in average precision.} Of note is that our model requires no hyperparameter optimization on some validation set, ensuring that it can be trained and prepared for use only using healthy data. 

\rev{Breast tumors can greatly vary in size between cases (Section \ref{sec:data}), so it is important that our model can detect both very small and very large tumors. We see in the case shown in the right top row of Fig. \ref{fig:comparemodels} that our method is able to localize an extremely small tumor, while other methods fail to do so. In the opposite case of a very large tumor shown in the right second row of the same figure, our model is also able to localize the tumor and make it stand out compared to the surrounding tissue area, despite the tumor being much larger than the size of the raster window/patch. This is a very important property of our method, as it allows for the localization of tumors of a wide range of sizes.}

\rev{We also note our model's performance on cases with dense breast tissue, shown in the bottom row of Fig. \ref{fig:comparemodels}, where the surrounding tissue of the tumor is visually similar to the tumor itself. As the tumor is not easily distinguishable from the surrounding tissue, this is a challenging case for both anomaly localization algorithms and radiologists \cite{nazari2018overviewdense}. However, our method is still able to localize the tumors in this case, while the other existing approaches both fail to differentiate it from its surroundings.}

\rev{Finally, we also evaluate the inference speed of our approach compared to existing methods, per image patch, shown in the rightmost column of Table \ref{tab:modelcompare}. We show that our method is about $1.3\times$ faster than CutPaste (0.062 sec. vs. 0.087 sec. per patch), and over $64\times$ faster than PatchSVDD (4.13 sec. per patch), while still possessing superior tumor detection performance. The large difference in inference speed between our method and PatchSVDD is due to the fact that PatchSVDD requires computing the distance of the patch's features to every single image's features in the training set, while our approach simply compares the patch's features to the features of the completions of the patch (in parallel).}

\subsection{Using State-of-the-Art Pluralistic Image Completion Backbones}
\label{sec:exp:compute}

As few research considers the topic of pluralistic image completion, we compare our dropout pluralistic completion method (Section \ref{sec:theory:dropout_inpaint}) to the state-of-the-art method of \cite{wan2021high}, which presents a two-stage, transformer-based model for pluralistic image completion.
We trained this method on the same random normal DBT patch dataset as the dropout inpainter, and example inpaintings created by the trained model are shown in Figure \ref{fig:eg_completions}. Although the completions are slightly more detailed (but still not anatomically valid) than our dropout inpainter, in practice we find that this method is \textit{significantly} slower than the dropout method for creating multiple completions, such that it becomes impractical for anomaly localization.

\begin{figure}
    \centering   
    \includegraphics[width=0.92\linewidth]{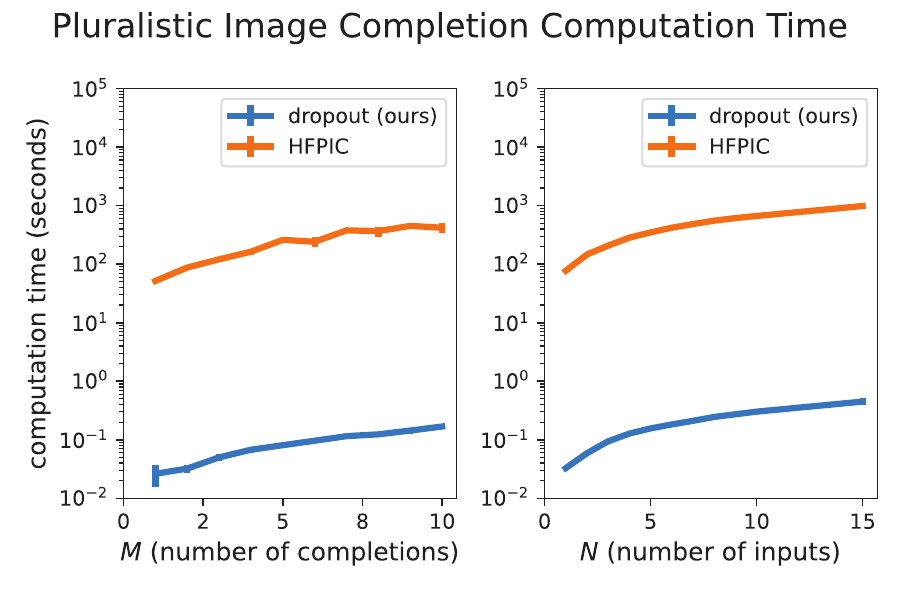}
    \caption{\textbf{Pluralistic image completion computational efficiency comparison.} All inpaintings completed with same $128\times 128$ center square mask, on a single RTX 3090 24 GB GPU. Note the logarithmic scale on the vertical (computation time) axis.}
    \label{fig:compute}
\end{figure}

On a single 24 GB RTX 3090 GPU, it takes 2.9 days for HFPIC and 8 minutes for our method to generate a single heatmap with the default setting of an $M=10$ completion sample size. This difference is simply due to the significant margin between the size of the two models: ours has about $3.6$ million trainable parameters, while HFPIC has about $450$ million. We further evaluated this difference by (1) fixing $N=5$, the number of input patches to complete, and testing a range of $M$ for both inpainting methods, and (2) fixing $M=10$ and testing a range of $N$. The computation time results are shown in Figure \ref{fig:compute}; each datapoint was averaged over six possible input DBT slice patches from the test set. We see that in general, HFPIC is slower than our method by about three orders of magnitude.

\begin{figure}
    \centering 
    \includegraphics[width=0.92\linewidth]{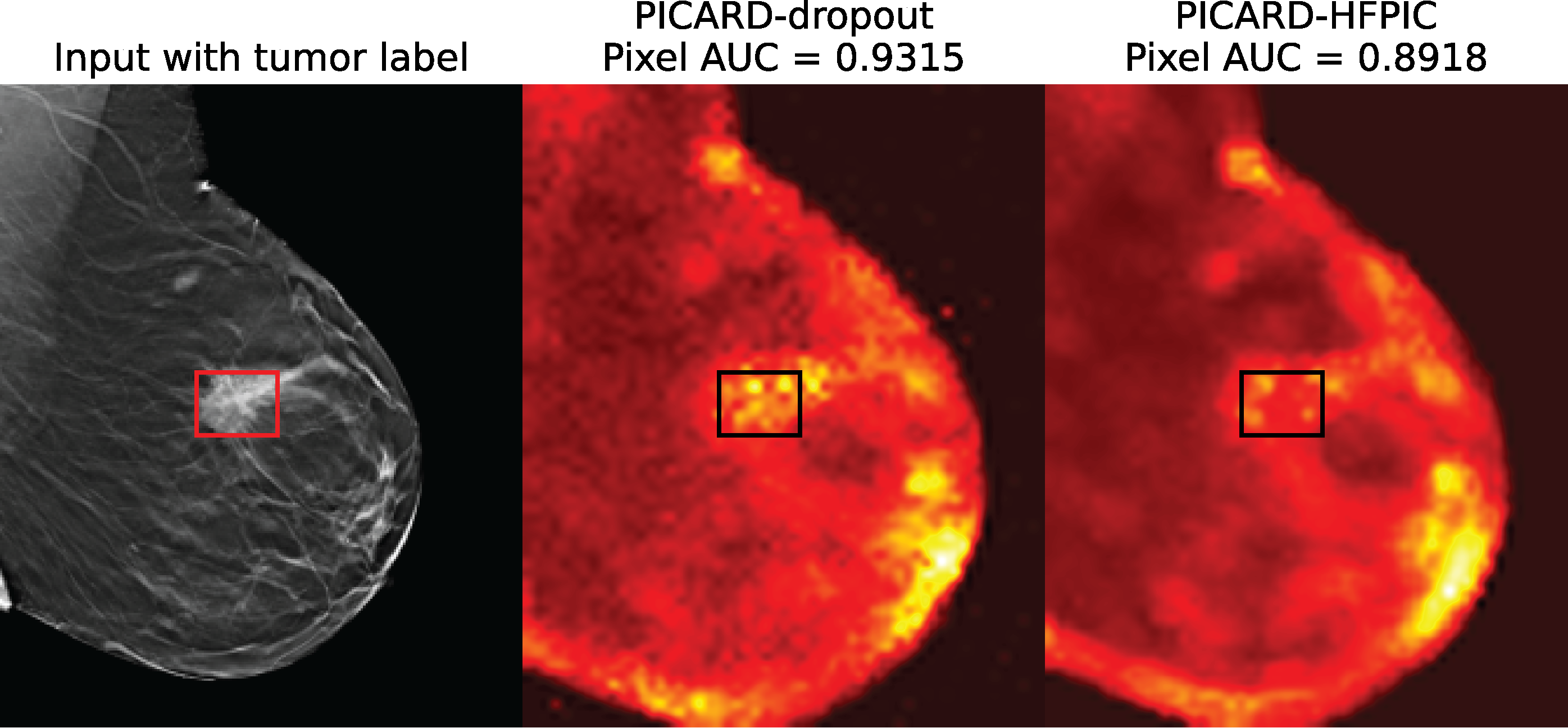}
    \caption{\textbf{PICARD heatmaps generated with different pluralistic image completion backbones.} From left to right: ground truth image with lesion label, and heatmaps generated using our dropout method, and HFPIC.}
    \label{fig:HFPICcompare}
\end{figure}

Moreover, we have tested the effectiveness of HFPIC by using it to create a heatmap for a DBT slice in the test set.
Although the extreme computation time makes it impractical to test PICARD with HFPIC on the entire test set, we tested it on a single image (which took days to compute) 
shown in Figure \ref{fig:HFPICcompare}. Here, we actually see a \textit{decrease} in anomaly localization performance; this is likely due to the anatomically unrealistic nature of HFPIC DBT completions, as shown in Figure \ref{fig:eg_completions}.

\subsection{Asymptotic Behavior of the MCD Metric}
\label{sec:exp:vsM}
\begin{figure}
    \centering    
    \includegraphics[width=0.8\linewidth]{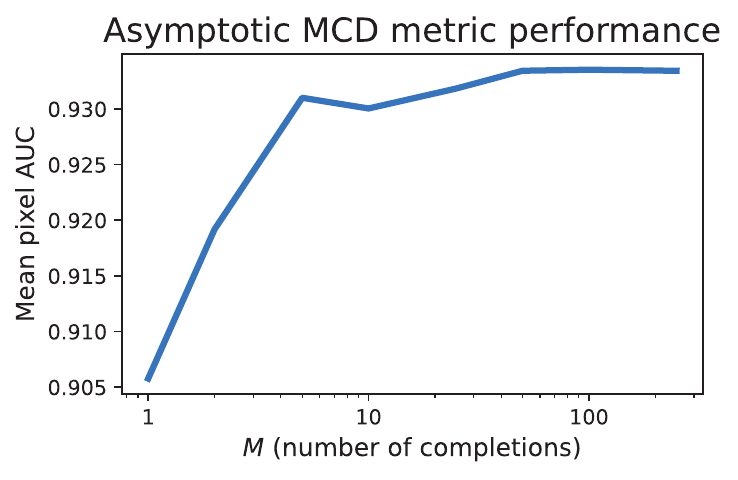}
    \caption{\textbf{Asymptotic anomaly localization performance using MCD metric}, with respect to number of completion samples $M$.}
    \label{fig:vsM}
\end{figure}

In Section \ref{sec:theory:auc_proof} we showed that theoretically, the MCD metric (Equation \ref{eq:score}) achieves perfect AUC performance in the limit of inpainting sample size $M\rightarrow\infty$. We evaluate this behavior empirically in order to validate these claims by calculating the tumor localization performance (pixel AUC) of PICARD on a range of values of $M$, $\{1, 2, 5, 10, 25, 50, 100, 250\}$, on a set of ten DBT scans randomly sampled from the test set, shown in Figure \ref{fig:vsM}. We use a subset instead of the full testing set due to computation feasibility (it takes almost 6 days to evaluate the entire set with $M=250$).  Performance does indeed increase asymptotically as $M\rightarrow\infty$, but not to a perfect AUC of $1$. This is due to the fact that in our derivations, we assume that the pluralistic inpainter is able to perfectly sample from the true distribution of possible completions; in practice, the inpainting method is necessarily imperfect, as it is difficult to capture the broad anatomical variability and complexity of breast tissue. Finally, we note that this analysis was completed after all other experiments, where $M=10$ was chosen \textit{a priori}.

\section{Discussion}
\label{sec:disc}
The central result of this work is that our pluralistic image completion-based anomaly localization (AL) method performs much better on DBT data than existing AL methods \cite{li2021cutpaste, patchsvdd} that have been shown to perform well on common machine learning AL benchmarks like MVTec-AD \cite{bergmann2019mvtec}. Importantly, these existing works differ from our approach in that they all rely on directly comparing the features of the input image to some learned distribution of normal features, not to \textit{new} normal (inpainting) features that are created by our model, and conditioned on the same surroundings as the input completion region. This fundamental difference in how anomaly localization/detection is approached is one reason for the superiority of our method. The MVTec-AD benchmark that the other methods do well with has normal data that vary minimally within a single object class, anomalous data that fall into one of several, in fact \textit{labeled} cases, and normal and anomalous data that have starkly different, and easily separable, features. These characteristics make anomaly detection easier, in terms of feature discrimination and generalization. However, DBT data does not possess these properties; healthy and cancer breast tissue possesses extreme semantic variability, and in many cases anomalous tissue can appear quite similar to healthy tissue (and vice versa). As such, it stands to reason that these existing methods generalize poorly when extended to DBT. We believe that this is excellent evidence for utilizing the \rev{BCS-DBT} dataset as a new benchmark for anomaly detection research in machine learning, due to the life-critical application yet high complexity of the data, \rev{and the fact that it is publicly available.}

\rev{While DBT tumor localization serves as a challenging benchmark for our anomaly localization algorithm, our approach is designed from a general standpoint, such that a wider range of applications are possible. As such, an important direction of future research is to extend our method to anomaly localization scenarios in other biomedical imaging modalities. These could include modalities such as OCT (optical coherence tomography), MRI (magnetic resonance imaging) or others.}

Interestingly, converting completions to the encoder feature space $\phi$ for PICARD did not introduce any performance boost as opposed to other AL methods; we hypothesize that this is again because breast tissue data is more complex and difficult to grasp useful features from than the natural image data that many of these other methods are built for. As we have already achieved strong performance with the current model, we leave it to future works to develop a feature encoder that could possibly be more robust to this type of data. Indeed, this could be related to the poor performance of the other, feature-discriminating AL methods, that do much better with natural or industrial images that have easier features to work with.

\subsection{Limitations}
The superior results of PICARD for DBT breast lesion detection are quite promising, however, further refinements could be applied to make the method even more powerful.

One of the difficulties of generative modeling of breast tissue is the extremely high complexity and natural variability of the tissue, so that it is difficult to obtain anatomically realistic completions, even with state-of-the-art methods like HFPIC \cite{wan2021high}. \rev{In addition to complex local details, breast tissue can have complicated correlations between distant image locations, which may not be able to be fully captured by the inherent locality of convolutional neural networks models. Indeed, the coarse-to-fine feature hierarchy of traditional convolutional image completion methods--such as the one used in this work--is well-suited for natural images, where low-resolution features are more global, yet it is unable to fully represent the complexities of breast tissue.} Visual transformer-based models, \eg, \cite{ViT,liu2021swin} can \rev{model long-range pixel interactions}, but even the transformer-based model of HFPIC was unable to produce anatomically realistic content. \rev{As such, it is unclear what type of generative model would be able to learn reasonable representations of breast data that preserve both local fine-grained details while maintaining the complex global structure of the tissue.} Such a model may need to include some sort of inductive biases for the unique structures seen in visual anatomical data; alternatively, entirely different generative models may prove useful, such as normalizing flows, energy-based, or score-based methods, which we leave for future works. 

Having more realistic completions would better approximate sampling from the true distribution of possible completions, theoretically leading to more robust minimum completion distance performance, and therefore better anomaly localization. This would fix some of the issues of false-positive regions that can be seen in some of PICARD's heatmaps, that have breast tissue that is labeled as healthy, but still possesses visual features that are uncommon in the training set. We found that even training our inpainter(s) on the full DBT training set of normal slices did not improve performance, so it appears to be a limitation of the model structure rather than the dataset size. 

Although PICARD's tumor localization performance does receive a boost from using multiple completions instead of just one (present in the first two rows of Table \ref{tab:modelcompare} where $M=10$, vs. the next two with $M=1$), in theory the difference could be higher, again if the sampled completions were more realistic and better approximated the true distribution of possible normal completions. One possible solution to this would be to choose a dropout probability for the completion network that results in optimal anatomical realism. However, such optimization needs access to some cancer images during the validation phase, greatly reducing its range of applications. It may be possible to quantify the anatomical realism of completions generated on some validation set of \textit{only} healthy cases, and optimize the dropout probability to maximize this quantity, but we leave this nontrivial task for future works.

Another possible future work is that as the completion region $I_c$ is our region of interest, we made no assumptions about if the surrounding region $I_m$ contains anomalies. Still, it may be worth considering how to detect anomalies within $I_m$ as well, which could begun with considering the joint distribution $p(I_c, I_m)$ rather than just $p(I_c|I_m)$ as in this work. However, it is unclear if this would improve heatmapping performance, as we use a stride small enough such that all pixels (beyond a ``padding region'' on the border) within a DBT slice will be included at least once within some evaluated $I_c$.

\section{Conclusion}
\label{sec:concl}
We introduced a novel anomaly localization method for ultra-high-resolution DBT breast scan data, called PICARD. We found that PICARD achieves promising performance with this difficult modality, that existing methods in the machine learning literature struggle to match. PICARD compares a distribution of \textit{pluralistic} normal image completions to the ground truth, and uses a new lightweight and efficient way to sample pluralistic completions using spatial dropout layers on a pretrained completion network. We also introduced a formal foundation for completion-based anomaly detection, and used it to mathematically analyze the convergence properties of our anomaly score. Finally, we synthesized all of these contributions into the final PICARD method.

\section*{Acknowledgments}
We would like to thank Jichen Yang and Brian Harrawood at Duke University for assisting with dataset downloading and management.

\section*{Funding}
This work was supported by Grant 1 R01 EB021360 from the National Institutes of Health (PI: Mazurowski).

{
    \small
    \bibliographystyle{ieeenat_fullname}
    \bibliography{main}
}

% SUPPLEMENTARY MATERIAL BELOW
% \clearpage
\appendix
\section*{Supplementary Material for ``Unsupervised anomaly localization in high-resolution breast scans using deep pluralistic image completion''}

\section{Implementational Details}
\label{app:implementation}

We implemented all of our methods and experiments in PyTorch. For all experiments, we fix the random seed to make the work reproducible. In all scripts we used a seed of $1337$ with the basic code shown below:
\begin{lstlisting}[language=Python]
import torch
import random

my_seed = 1337

random.seed(my_seed)
torch.manual_seed(my_seed)
torch.cuda.manual_seed_all(my_seed)
\end{lstlisting}

\subsection{Network Architecture and Experimental Details for PICARD (Our Method)}
\label{app:exp_ours}
When applying spatial dropout to the (convolutional) layers of $G$, we also found it important to not just apply dropout to every single layer, but to exclude certain layers from dropout for better performance.
In particular we found it essential to not apply dropout to the first or last convolutional layers of $G_{fine}$ (\texttt{conv1} and \texttt{conv17}) and $G_{coarse}$ (\texttt{conv1} and \texttt{allconv17}), as not doing so can result in possible identical ``noise'' completions that are occasionally sampled with a frequency proportional to the dropout probability. For this particular completion network we also found that excluding dropout on all of the layers following the final atrous/dilated layer \texttt{conv10\_atrous} improved the quality of completions. Specifically, we observed that if these layers \textit{are} included for dropout, occasionally dark regions appeared in feature maps on this part of the network that seeded unrealistic regions in the final completion. We reason that this is because these post-atrous layers perform interpolation to upsample the feature maps to increase the resolution to the final output resolution, so that initially small artifacts may become much more problematic as data is passed through further layers.

To convert completions in image space to a useful feature space, we use the feature map output of the final convolutional layer (\texttt{conv4}) of the WGAN critic as our encoder $\phi$, flattened to be in $\mathbb{R}^{8192}$. We also note that this critic $\phi$ is actually the \textit{local} critic in the original work of \cite{yu2018generative}, takes completion regions as inputs alone. This is opposed to the \textit{global} critic, which takes entire completed images as inputs. We pretrained the completion network $G$ and critic/encoder $\phi$ on the DBT patch dataset using the default procedure and hyperparameters of \cite{yu2018generative}, with a batch size of $55$. We trained until we saw the $L_1$ reconstruction error between the inpaintings and the ground truths get no lower, at 130,000 iterations, on two NVIDIA RTX 3090 24GB GPUs. We generated all PICARD heatmaps with four 3090 24GB GPUs.

We also experimented with only including pixels within the breast (not in the outside black region that is within each slice) for the AUC calculation/heatmapping procedure, but we decided that it would be best to include the entire image to have the least number of experimental biases as possible, and to observe the performance of all models in \textit{all} parts of the slice.

\subsection{Experimental Details for Other Methods}
\label{app:exp_others}
\paragraph{HFPIC \cite{wan2021high}}

\textbf{Transformer/coarse prior generator training:} 
We used the default setting that the HFPIC authors used for ImageNet \cite{russakovsky2015imagenet}, except trained from scratch on the BCS-DBT training set (Table \ref{tab:data}). Specifically, we used the BERT training objective, the \texttt{GELU\_2} activation function, and randomly generated \texttt{pconv} completion masks. For the transformer, we used 35 layers, an embedding size of 1024, and 8 heads. We trained for 200 epochs on a batch size of 6.

\textbf{Convolutional network/guided upsampler training:} 
Just as for the transformer, we used the default setting that the HFPIC authors used for ImageNet, except trained from scratch on the BCS-DBT training set (Table \ref{tab:data}). We trained with a batch size of $75$ for $40,000$ iterations, on randomly generated \texttt{pconv} completion masks.

We chose the best performing models from each of their training according to the paper's original validation score on normal image patches. Inference was performed with the same parameters as in training. We performed all experiments on one 48 GB NVIDIA RTX A6000 GPU.

\subsection{Computational Efficiency}
\label{app:compute}
Our heatmapping model enjoys high scalability, by allowing for both the computation of multiple completions for a given input, and the computation of multiple inputs, all at the same time. This was created using PyTorch's inherent support for parallelism; the former is completed by inputting $M$ copies of the same input image to the image completion model as a single, parallelized batch, while the latter is completed by creating a custom data loader that can load batches of heatmap raster windows, to be analyzed all at once.

One heatmap for a full size ($\sim2,000\times2,500$) DBT image is generated in just two minutes by four RTX 3090 GPUs. This is achievable by virtue of the relative simplicity and lack of additional computational load created by our dropout-based multi-inpainting method.

\section{\rev{Ablation Studies}}
\label{app:exp}
\rev{
\subsection{Mean or Median Completion Distance instead of Minimum}
By default, our approach uses the minimum completion distance/MCD anomaly scoring metric (equation \eqref{eq:score}). Here we compare the tumor detection performance of the MCD metric with anomaly metrics that use the \textit{mean} or \textit{median} distance of normal completion samples to the ground truth,
\begin{equation}
\label{eq:score_mean}
\mathcal{A}^{mean}_M\left(I_c;I_m\right)\triangleq\frac{1}{M}\sum\limits_{h_c^i\sim p_n\left(h_c\middle|I_m\right)}\left|\left|h_c^0-h_c^i\right|\right|_2
\end{equation}
and
\begin{equation}
    \label{eq:score_median}
    \mathcal{A}^{median}_M\left(I_c;I_m\right)\triangleq\mathop{\mathrm{median}}\limits_{h_c^i\sim p_n\left(h_c\middle|I_m\right)}\left\{\left|\left|h_c^0-h_c^i\right|\right|_2\right\},
\end{equation}
respectively. Evaluated on tumor detection for the DBT test set as in Section \ref{sec:exp:comparison}, we found that our MCD metric outperforms the mean and median metrics, as shown in Table \ref{tab:app_exp:mean_median}.
}

\begin{table}
    \centering
    \caption{Quantitative comparison of using the minimum (default, equation \ref{eq:score}), mean (equation \ref{eq:score_mean}) and median (equation \ref{eq:score_median}) completion distance anomaly score metrics for PICARD for tumor localization on the DBT test set.}
    \begin{tabular}{lc}
        \toprule
        Anomaly Metric & Pixel AUC \\
        \midrule
        Min. CD/MCD \textit{(image space)} & \textbf{0.875} \\
        Min. CD/MCD \textit{(feature space)} & 0.865 \\
        Mean CD \textit{(image space)} & 0.863 \\
        Mean CD \textit{(feature space)} & 0.843 \\
        Median CD \textit{(image space)} & 0.867 \\
        Median CD \textit{(feature space)} & 0.846 \\
        \bottomrule
    \end{tabular}
    \label{tab:app_exp:mean_median}
\end{table}

\subsection{Regular Dropout Instead of Spatial Dropout}
In practice, we found that using regular dropout on $G$ can introduce pixelated artifacts within completions and/or significantly less plausible completions than for when we used spatial dropout.

\section{Dataset Details}
\label{app:data}
In this paper 
we work with the BCS-DBT (Breast Cancer Screening-Digital Breast Tomosynthesis) dataset of ultra-high-resolution 3D breast cancer scans, originally from \cite{buda2020detection} and recently published as \cite{buda2021data}. This dataset is a collection of $22,032$ breast scans that are divided into four disjoint classes, of which we use \textit{normal} (no potential lesions flagged by a radiologist), and \textit{cancer} (contains lesion(s) flagged by a radiologist, biopsied and confirmed to be cancer). Each scan is a 3D \textit{volume} of about 70 physically-adjacent 2D greyscale scan \textit{slices}, where each slice is a $2457\times1890$ image. For our training set we take $6,245$ healthy slices from the BCS-DBT training set, from $2,000$ different patients. Each slice is sampled randomly from a different anatomical view/volume, originating from $2,000$ different patients in total. For our test set, we take all $133$ slices from the BCS-DBT test set that have radiologist-annotated biopsied lesion(s) (as each lesion annotation corresponds to one of the slices, or in rare cases, multiple lesions per slice). The massive class imbalance is due to the rarity of breast cancer: for regular screening mammograms, only about $0.6\%$ of women obtain positive result for cancer \cite{lehman2017national}.

The BCS-DBT dataset is useful for evaluating image anomaly detection methods for a number of reasons:
\begin{enumerate}
    \item It represents an important real-world application of anomaly detection: breast cancer detection, which is a leading cause of death in women. Approximately 1 in 8 women will be diagnosed with invasive breast cancer in their lifetime, and 1 in 39 women will die from it \cite{howlader1975seercancer}. On the other hand, many anomaly detection benchmarks seen in the general machine learning literature \eg, \textit{CIFAR-10} and \textit{CIFAR-100} do not clearly represent real-world anomaly detection use cases. The industrial anomaly dataset MVTec \cite{bergmann2019mvtec} is an important exception to this, but it is still considerably simpler than medical anomaly datasets.
    \item The data is very high resolution, both normal and anomalous data have strong semantic variability within the two classes, and anomalous data can often appear similar to normal data, altogether forming a challenging dataset.
\end{enumerate}
Furthermore, the dataset is specifically useful for our method because (1) there are an abundance of normal instances, which is necessary to train GAN based models, especially for high-resolution data; (2) there are radiologist-annotated bounding boxes that provide ground truths of cancerous lesions/anomalous data. Other anomaly detection benchmarks either do not possess bounding boxes for object instances, like \textit{CIFAR-10}, \textit{CIFAR-100} \cite{cifar}, \textit{FashionMNIST} \cite{xiao2017fashion} and \textit{CatsVsDogs} \cite{catsvdogs}, and/or do not have enough per-class instances to successfully train our GAN-based method on, such as \textit{102 Category Flowers} \cite{flowers}, \textit{Caltech UCSD Birds 200} \cite{wah2011caltech}, \textit{MVTec} \cite{bergmann2019mvtec}, \textit{WBC} \cite{wbc}, or \textit{DIOR} \cite{dior}. The first criteria is important because our method is defined by the conditional information of the surroundings of the region of interest, which needs a bounding box that is known whether or not an anomalous object is within it. Still, in a future work we would like to attempt using a completion model that can be trained on low amounts of data, so that these datasets with relatively few per-class instances that \textit{do} have anomaly/object bounding boxes, \eg, \textit{MVTec}, could be tested on our method.

For our dropout inpainter, we pretrain the completion network $G$ and the critic/encoder $\phi$ on random $256\times 256$ patches of the training set. The network is trained to inpaint random rectangular masks of normal patches, while for testing, we always use $128\times128$ centered square masks. We normalize all data to the range $[-1, 1]$ as used in \cite{yu2018generative}. We train the HFPIC inpainter on the same dataset of healthy scan patches.

The vast majority of the BCS-DBT dataset is publicly available (on \url{https://wiki.cancerimagingarchive.net/pages/viewpage.action?pageId=64685580} ), except for a small portion, described as follows. The dataset is originally divided into a training set ($19,148$ DBT scan volumes/DICOM files), a validation set ($1,163$) and a test set ($1,721$). All raw image/DICOM files are public. Each of these sets have normal and cancer class instances, with two other classes that are not relevant for this work. The training set has a public list of both class labels for all scans, and lesion bounding boxes for cancerous scans. 

However, the BCS-DBT validation and test sets do not currently include public labels and bounding boxes, which we plan to release soon. 

Now, our method could still be trained and tested on the labeled data that IS publicly available, namely BCS-DBT's training set. This training set contains $18,232$ total DBT volumes, a subset of which we used for training, and it contains 76 cancer scans that we did not use, that could be used for testing.

% WARNING: do not forget to delete the supplementary pages from your submission 
% \input{sec/X_suppl}

\end{document}